\documentclass[twocolumn,showpacs,preprintnumbers,amsmath,amssymb,prd,floatfix]{revtex4}

\usepackage{graphicx}
\usepackage{dcolumn}
\usepackage{bm}

\renewcommand{\rho}{\varrho}

\newcommand{\mone}{\multicolumn{1}{c}}

\newcommand{\mum}{\ensuremath \mu\mbox{m}}
\newcommand{\mug}{\ensuremath \mu\mbox{g}}
\newcommand{\Gunit}{\ensuremath \mbox{m}^3 \;\mbox{kg}^{-1}\;\mbox{s}^{-2}}

\newcolumntype{q}{D{.}{.}{1}}
\newcolumntype{s}{D{.}{.}{2}}
\newcolumntype{t}{D{.}{.}{3}}
\newcolumntype{u}{D{.}{.}{4}}
\newcolumntype{v}{D{.}{.}{5}}
\newcolumntype{w}{D{.}{.}{6}}
\newcolumntype{x}{D{.}{.}{7}}
\newcolumntype{y}{D{.}{.}{9}}

\begin{document}

\title{A Measurement of Newton's Gravitational Constant}
\author{St. Schlamminger}
\altaffiliation[present address ]{Univ. of Washington, Seattle, Washington, USA}
\author{E. Holzschuh}
\altaffiliation{deceased}
\author{W. K\"undig}
\altaffiliation{deceased} \altaffiliation{We dedicate this paper
to our colleague Walter K\"{u}ndig, without whose untiring and
persistent effort this ambitious experiment would neither have
been started nor brought to a successful conclusion. Walter
K\"{u}ndig, died unexpectedly and prematurely of a grave illness
in May 2005. He conceived the set-up of this experiment and worked
on aspects of the analysis until a few days before his death.}

\author{F. Nolting}
\altaffiliation{Paul Scherrer Institut,Villigen, Switzerland}
\author{R.E. Pixley}
\email[EMail address: ]{ralph.pixley@freesurf.ch}
\author{J. Schurr}
\altaffiliation[present address ]{Physikalisch Technische Bundesanstalt, Braunschweig, Germany}
\author{U. Straumann}
\affiliation{Physik-Institut der Universit\"{a}t Z\"{u}rich, CH-8057 Z\"{u}rich, Switzerland}

\begin{abstract}
A precision measurement of the gravitational constant $G$ has been
made using a beam balance. Special attention has been given to
determining the calibration, the effect of a possible nonlinearity
of the balance and the zero-point variation of the balance. The
equipment, the measurements and the analysis are described in
detail. The value obtained for $G$ is 6.674252(109)(54) $\times
10^{-11}$\;$\Gunit$. The relative statistical and systematic
uncertainties of this result are 16.3$\times 10^{-6}$ and
8.1~$\times 10^{-6}$, respectively.
\end{abstract}


\maketitle

\section{Introduction}
The gravitational constant $G$ has proved to be a very difficult
quantity for experimenters to measure accurately. In 1998, the
Committee on Data Science and Technology (CODATA) recommended a
value of $6.673(10)\times 10^{-11}\; \Gunit$. Surprisingly,  the
uncertainty, 1,500 parts per million (ppm), had been increased by
a factor of 12 over the previously adjusted value of 1986. This
was due to the fact that no explanation had been found for the
large differences obtained in the presumably more accurate
measurements carried out since 1986. Obviously, the differences
were due to very large systematic errors. The most recent revision
\cite{Mo05} of the CODATA Task Group gives for the 2002
recommended value $G=6.6742(10)\times 10^{-11}\; \Gunit$. The
uncertainty (150 ppm) has been reduced by a factor of 10 from the
1998 value, but the agreement among the measured values considered
in this compilation is still somewhat worse than quoted
uncertainties.

\begin{figure}
\includegraphics[height=10cm]{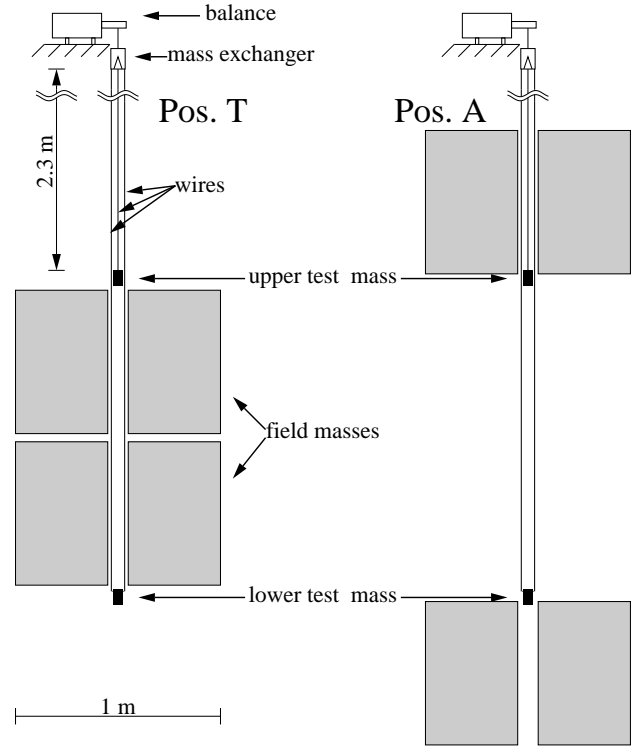}
\caption{Principle of the measurement. The FM's are shown in the
position together (Pos. T) and the position apart (Pos. A).}
\label{fig1}
\end{figure}

Initial interest in the gravitational constant at our institute
had been motivated by reports \cite{Fi86} suggesting the existence
of a ``fifth" force which was thought to be important at large
distances. This prompted  measurements at a Swiss storage lake in
which the water level varied by 44~m. The experiment involved
weighing two test masses (TM's) suspended next to the lake at
different heights. No evidence \cite{Co94,Hu95} was found for the
proposed ``fifth" force, but, considering the large distances
involved, a reasonably accurate value (750~ppm) was obtained for
$G$. It was realized that the same type of measurement could be
made in the laboratory with much better accuracy with the lake
being replaced by the well defined geometry of a vessel containing
a dense liquid such as mercury. Equipment for this purpose was
designed and constructed in which two 1.1~kg TM's were alternately
weighed in the presence of two moveable field masses (FM's) each
with a mass of 7.5~t. A first series  of measurements
\cite{Sa98,Sb98,No98,No99,Sa99} with this equipment resulted in a
value for $G$ with an uncertainty of 220~ppm due primarily to a
possible nonlinearity of the balance response function. A second
series of measurements was undertaken to eliminate this problem. A
brief report of this latter series of measurements has been given
in ref. \cite{Sa02} and a more detailed description in a thesis
\cite{Sb02}. Since terminating the measurements, the following
four years have been spent in improving the analysis and checking
for possible systematic errors.

Following a brief overview of the experiment, the measurement and
the analysis of the data are presented in  Secs.~\ref{sec3} entitled
Measurement of the Gravitational Signal and Sec.~\ref{sec4} entitled
Determination of the Mass-Integration Constant. In Sec.~\ref{sec5}, the
present result is discussed and compared with other recent
measurements of the gravitational constant.

\section{General Considerations} The design goal of this
experiment was that the uncertainty in the measured value of $G$
should be less than about 20~ppm. This is comparable to the quoted
accuracy of recent $G$ measurements made with a torsion balance.
It is, however, several orders of magnitude better than previous
measurements of the gravitational constant (made after 1898)
employing a beam balance \cite{Hu95,Mo88,Sp87}.

The experimental setup is illustrated in Fig.~\ref{fig1}. Two nearly
identical 1.1~kg TM's hanging on long wires are alternately
weighed on a beam balance in the presence of the two movable FM's
weighing 7.5~t each. The position of the FM's relative to the TM's
influence the measured weights. The geometry is such that when the
FM are in the position labelled "together", the weight of the
upper TM is increased and that of the lower TM is decreased. The
opposite change in the TM weights occurs when the FM are in the
position labelled "apart". One measures the difference of TM
weights first with one position of the FM's and then with the
other. The difference between the TM weight differences for the
two FM positions is the gravitational signal.

The use of two TM's and two FM's has several advantages over a
single TM and a single movable FM. Comparing two nearly equivalent
TM's tends to cancel slow variations such as zero-point drift of
the balance and the effect of tidal variations. Using the
difference of the two TM weights doubles the signal. In addition,
it causes the influence of the FM motion on the counter weight of
the balance to be completely cancelled. Use of two FM's with equal
and opposite motion reduces the power required to that of
overcoming friction. This also simplified somewhat the mechanical
construction.

The geometry has been designed such that the TM being weighed is
positioned at (or near) an extremum  of the vertical force field
in both the vertical and horizontal directions for both positions
of the FM's. The extremum is a maximum for the vertical position
and a minimum for the horizontal position. This double extremum
greatly reduces the positional accuracy required in the present
experiment.

\begin{figure}
\includegraphics[width=8.5cm]{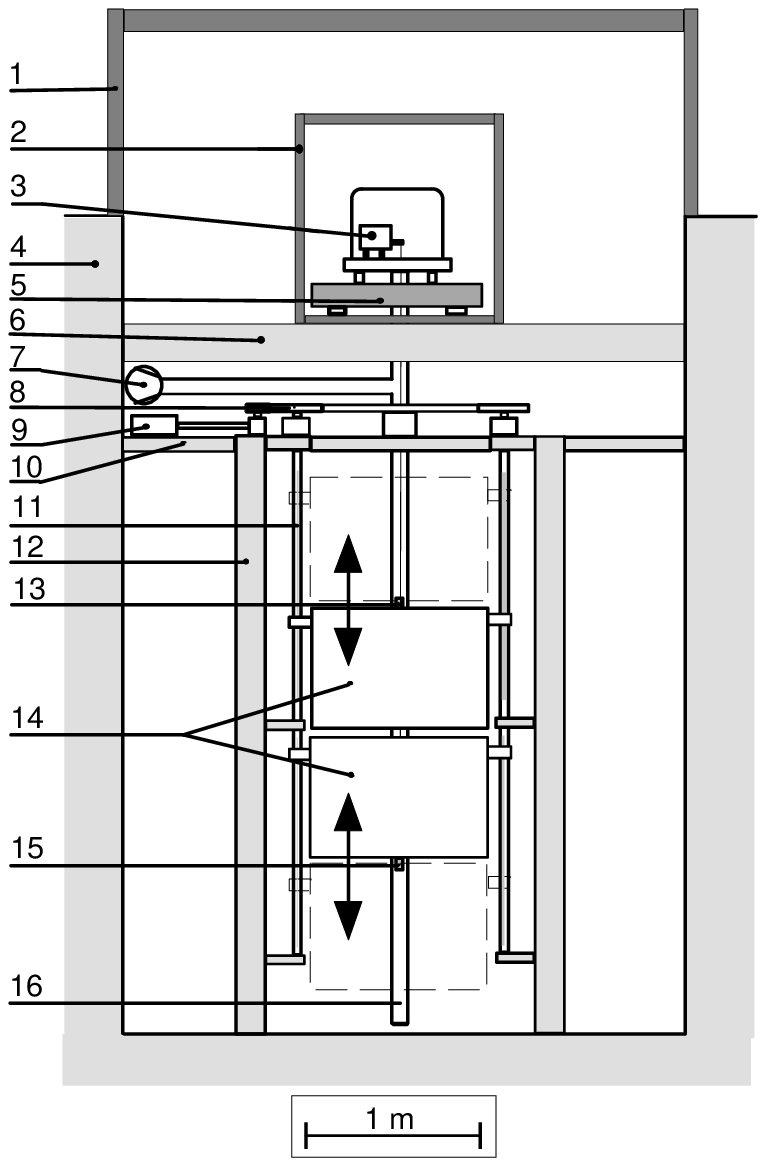}
\caption[A side view of the experiment.]{A side view of the
experiment. Legend: 1=measuring room enclosure, 2=thermally
insulated chamber, 3=balance, 4=concrete walls of the pit,
5=granite plate, 6=steel girder, 7=vacuum pumps, 8=gear drive,
9=motor, 10=working platform, 11=spindle, 12=steel girder of the
main support, 13=upper TM, 14=FM's, 15=lower TM, 16=vacuum tube.}
\label{fig2}
\end{figure}

The measurement took place at the Paul Scherrer Institut (PSI) in
Villigen. The apparatus was installed in a pit with thick concrete
walls which provided good thermal stability and isolation from
vibrations. The arrangement of the equipment is shown in
Fig.~\ref{fig2}. The system involving the FM's was supported by a
rigid steel structure mounted on the floor of the pit. Steel
girders fastened to the walls of the pit supported the balance,
the massive (200 kg) granite plate employed to reduce high
frequency vibrations and the vacuum system enclosing the balance
and the TM's. A vacuum of better than $10^{-4}$~Pa was produced by
a turbomolecular pump located at a distance of 2~m from the
balance.

The pit was divided into an upper and a lower room separated by a
working platform 3.5~m above the floor of the pit. All heat
producing electrical equipment was located in the upper measuring
room. Both rooms had their own separate temperature stabilizing
systems. The long term temperature stability in both rooms was
better than $0.1^\circ$~C. No one was allowed in either room
during the measurements in order to avoid perturbing effects.

The equipment was fully automated. Measurements lasting up to 7
weeks were essentially unattended. The experiment was controlled
from our Zurich office via the internet with data transfer
occurring once a day.

\section{Measurement of the Gravitational Signal}
\label{sec3}
We begin this section with a description of the devices employed
in determining the gravitational signal. Following the
descriptions of these devices, the detailed schedule of the
various weighings and their analysis are given. Balance weighings
will be expressed in mass units rather than force units. The value
of local gravity was determined for us by E. E. Klingel\'{e} of
the geology department of the Swiss Federal Institute. The
measurement was made near the balance on Sept. 11, 1996 using a
commercial gravimeter (model G \#317 made by the company
LaCoste-Romberg). The value found was $9.807 233 5(6)\;\mbox{m}\mbox{s}^{-2}$.
This value was used to convert the balance readings into force
units.

\subsection{The Balance}
\label{sec31} The beam balance was a modified commercial mass
comparator of the type AT1006 produced by the Mettler-Toledo
company. The mass being measured is compensated by a counter
weight and a small magnetic force between a permanent magnet and
the current flowing in a coil mounted on the balance arm. An
opto-electrical feedback system controlling the coil current
maintains the balance arm in essentially a fixed position
independent of the mass being weighed. The digitized coil current
is used as the output reading of the balance.

The balance arm is supported by two flexure strips which act as
the pivot. The pan of the balance is supported by a parallelogram
guide attached to the balance frame. This guide allows only
vertical motion of the pan to be coupled to the arm of the
balance. Horizontal forces produced by the load are transmitted to
the frame and have almost no influence on the arm.

As supplied by the manufacturer, the balance had a measuring range
of 24~g above the 1~kg offset  determined by the counter weight.
The original readout resolution was 1~\mug\ and the specified
reproducibility was 2~\mug. The balance was designed especially
for weighing a 1 kg standard mass such as is maintained in many
national metrology institutes.

In the present experiment, the balance was modified by removing
some nonessential parts of the balance pan which resulted in its
weighing range being centered on 1.1~kg instead of the 1~kg of a
standard mass. Therefore, 1.1~kg TM's were employed. In order to
obtain higher sensitivity required for measuring the approximately
0.8~mg difference between TM weighings, the number of turns on the
coil was reduced by a factor of 6, thus reducing the range to 4~g
for the same maximum coil current. The balance was operated at an
output value near 0.6~g which gave a good signal-to-noise ratio
with low internal heating. For the present measurements, a mass
range of only 0.2~g was required. The full readout resolution of
the analog to digital converter (ADC) measuring the coil current
was employed which resulted in a readout-mass resolution of
12.5~ng.

An 8th order low-pass, digital filter with various time constants
was available on the balance. Due to the many weighings required
by the procedure employed to cancel nonlinearity (see
Sec.~\ref{sec34}), it was advantageous to make the time taken for
each weighing as short as possible. Therefore, the shortest filter
time constant (approximately 7.8~s) was employed and output
readings were taken at the maximum repetition rate allowed by the
balance (about 0.38~s between readings).

Pendulum oscillations were excited by the TM exchanges. Small
oscillation amplitudes (less than 0.2~mm) of the TM's
corresponding to one and two times the frequency for pendulum
oscillations (approximately 0.26~Hz for the lower TM and 0.33~Hz
for the upper TM's) were observed. They were essentially undamped
with decay times of several days. The unwanted output amplitudes
of these pendulum oscillation were not strongly attenuated by the
filter (half-power frequency of 0.13~Hz) and therefore had to be
taken into account in determining the equilibrium value of a
weighing.

The equilibrium value of a weighing was determined in an on-line,
5-parameter, linear least-squares fit made to 103 consecutive
readings of the balance starting 40~s after a load change. The
parameters of the fit were 2 sine amplitudes, 2 cosine amplitudes
and the average weight. The pendulum frequencies were known from
other measurements and were not parameters of the on-line fit. The
40~s delay before beginning data taking was required in order to
allow the balance to reach its equilibrium value (except for
oscillations) after a load change. This procedure (including the
40~s wait) is what we call a "weighing". A weighing thus required
about 80~s.

\begin{figure}
\includegraphics[width=8.5cm]{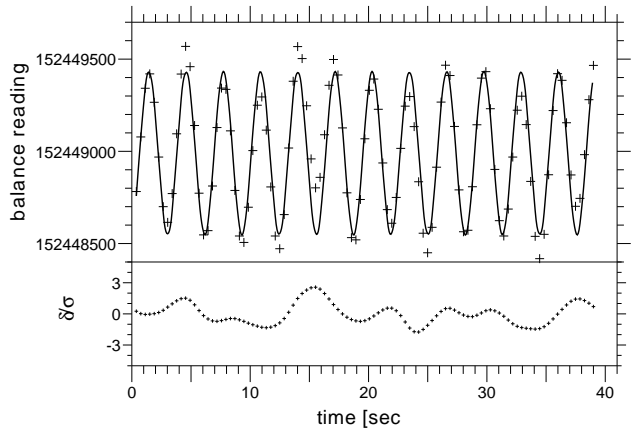}
\caption{Shown in the upper plot are the balance readings for a
typical weighing illustrating the oscillatory signal due to
pendulum oscillations. The output is the uncalibrated balance
reading corresponding to approximately 1.1 kg with a magnetic
compensation of 0.6 g. The amplitude of the oscillatory signal
corresponds to about 1.5 \mug. The lower plot shows the normalized
residuals. The normalization has been chosen such that their rms
value is 1.} \label{fig3}
\end{figure}

Data of a typical weighing and the fit function used to describe
their time distribution are shown in the upper part of
Fig.~\ref{fig3}. The residuals $\delta$ divided by a normalization
constant $\sigma$ are shown in the lower part of this figure. The
normalization constant has been chosen such that the rms value of
the residuals is 1. Since the balance readings are correlated due
to the action of the digital filter, the value of $\sigma$ does
not represent the uncertainty of the readings. It is seen that the
residuals show only rather wide peaks. These peaks are probably
due to very short random bursts of electronic noise which have
been widened by the digital filter. With the sensitivity of our
modified balance, they represent a sizable contribution to the
statistical variations of the weighings. They are of no importance
for the normal use of the AT1006 balance.

A direct calibration of the balance in the range of the 780 \mug\
gravitation signal can not be made with the accuracy required
in the present experiment ($< 20$~ppm) since calibration masses of
this size are not available with an absolute accuracy of better
than about 300~ppm. Instead, we have employed a method in which an
accurate, coarse grain calibration was made using two 0.1 g
calibration masses (CM's). The CM's were each known with an
absolute accuracy of 4~ppm. A number of auxiliary masses (AM's)
having approximate weights of either 783 $\mu g$ or $16\times
783=12,528$~\mug\ were weighed along with each TM in steps of
783~\mug\ covering the 0.2~g range of the CM. Although the AM's
were known with an absolute accuracy of only 800~ng (relative
uncertainty 1,000~ppm), the method allowed balance nonlinearity
effects to be almost entirely cancelled. Thus, the effective
calibration accuracy for the average of the TM difference
measurements was essentially that of the CM's. A detailed
description of this method is given in Sec.~\ref{sec310}.

In our measurements, the balance was operated in vacuum. The
balance proved to be extremely temperature sensitive which was
exacerbated by the lack of convection cooling in vacuum. The
measured zero-point drift was 5.5~mg/$^\circ \mbox{C}$. The sensitivity
of the balance changed by 220~ppm$/^\circ \mbox{C}$. To reduce these
effects, the air temperature of the room was stabilized to about
0.1~$^\circ \mbox{C}$. A second stabilized region near the balance was
maintained at a constant temperature to 0.01~$^\circ \mbox{C}$. Inside
the vacuum, the balance was surrounded by a massive (45~kg) copper
box which resulted in a temperature stability of about 1~mK.
Although zero-point drift under constant load for a 1~mK
temperature change was only 5.5~\mug\ , the effects of self
heating of the balance due to load changes during the measurement
of the gravitational signal were much larger. Details of this
effect and how they were corrected are described in Sec.\ref{sec37}.

\subsection{The Test Masses}
\label{sec32} One series of measurements was made using copper
TM's and two with tantalum TM's. Various problems with the mass
handler occurred during the measurements with the tantalum TM's
which resulted in large systematic errors. Although the tantalum
results were consistent with the measurements with the copper
TM's, the large systematic errors resulted in large total errors.
The tantalum measurements were included in our first publication,
but we now believe that better accuracy is obtained overall with
the copper measurements alone. We therefore describe only the
measurements made with the copper TM's in the present work.

\begin{figure}
\includegraphics[width=8.5cm]{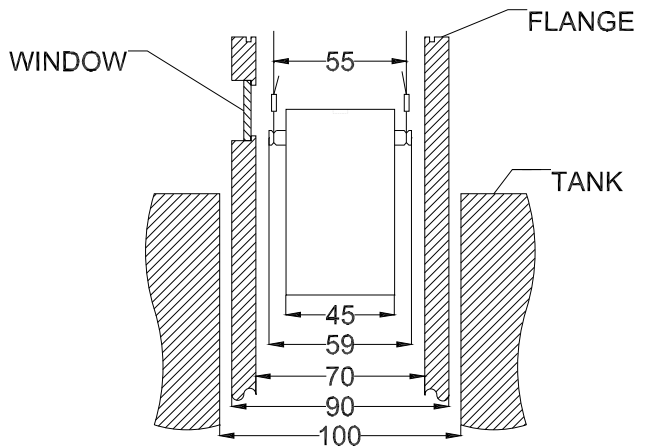}
\caption{Drawing of TM inside the vacuum tube. Dimensions are
given in mm. } \label{fig:cotm}
\end{figure}

A drawing of a copper TM is shown in Fig.~\ref{fig:cotm}. The
45~mm diameter, 77 mm high copper cylinders were plated with a
10~\mum\ gold layer to avoid oxidation. The gold plating was made
without the use of nickel in order to avoid magnetic effects. Near
the top of each TM on opposite sides of the cylinder were two
short horizontal posts. The posts were made of Cu-Be (Berylco 25).
The tungsten wires used to attach the TM's to the balance were
looped around these posts in grooves provided for this purpose.
The wires had a diameter of 0.1~mm and lengths of 2.3~m for the
upper TM and 3.7~m for the lower. The loop was made by crimping
the tungsten wire together in a thin copper tube. A thin,
accurately machined, copper washer was placed in a cylindrical
indentation on the top surface of the lower TM in order to trim
its weight (including suspension) to within about 400~\mug\ of
that of the upper TM and suspension.

Measurement of the TM's dimensions was made with an accuracy of
5~\mum\ using the coordinate measuring machine (CMM) at PSI.  The
weight of the gold plating was determined from the specified
thickness of the layer. The weight of the tungsten wires was
determined from the dimensions and the density of tungsten. The
thin tubing used to crimp the tungsten wires was weighed directly.
The weight of the complete TM's was determined at the
Mettler-Toledo laboratory with an accuracy of 25~\mug\ (0.022~ppm)
before and after the gravitational measurement. It was found that
the mass of both TM's had increased by a negligible amount
(0.5~ppm) during the measurement.

An estimate of possible density gradients in the TM's was
determined by measuring the density of copper samples bordering
the material used for making the TM's. It was found that the
variation of the relative density gradients over the dimensions of
either TM was less than $2\times 10^{-4}$ in both the longitudinal
and the radial directions.

\subsection{TM Exchanger}
\label{sec33} In weighing the TM's, it was necessary to remove the
suspension supporting one TM from the balance and replace it by
the other supporting the other TM. The exchange was accomplished
by a step-motor driven hydraulic systems to raise the suspension
of one TM while lowering the other. A piezo-electric transducer
mounted above the pan of the balance was used to keep the load on
the balance during the exchange as constant as possible. This was
done in order to avoid excessive heating due to the coil current
and to reduce anelastic effects in the flexure strips supporting
the balance arm. The output excursions were typically less than
0.1~g. The exchange of the TM's required about 4~min.

The TM suspension rested on a thin metal arm designed to bend
through 0.6~mm when loaded with 1.1~kg. Therefore, the transfer of
TM's was accomplished with a vertical movement of typically 2~mm
(0.6~mm bending of the spring plus an additional 1.4~mm to avoid
electrostatic forces). The metal arm was attached to a
parallelogram guide (similar to that of the balance) to assure
only vertical motion.

Although the parallelogram reduced the error resulting from the
positioning of the load, it was nevertheless important to have the
TM load always suspended from the same point on the balance pan.
This was accomplished by means of a kinematic coupling
\cite{Sl88,Fu81}. The coupling consisted of three pointed titanium
pins attached to each TM suspension which would come to rest in
three titanium V grooves mounted on the balance pan. The
reproducibility of this positioning was 10~\mum. The pieces of the
coupling were coated with tungsten carbide to avoid electrical
charging and reduce friction.

\subsection{Auxiliary Masses}
\label{sec34} In order to correct for any nonlinearity of the
balance in the range of the signal, use was made of many auxiliary
masses (AM's) spanning the 200~mg range of the CM's in steps of
approximately 783~\mug. Although the AM's could not be measured
with sufficient accuracy to calibrate the balance absolutely, they
were accurate enough to correct the measured gravitational signal
for a possible nonlinearity of the balance. Each TM was weighed
along with various combinations of AM's. One essentially averaged
the nonlinearity over the 200 mg range of the CM's in 256 load
steps of 200~mg/256=783~\mug. A weighing of both 100~mg CM's was
then used to determine the absolute calibration of the balance
which is valid for the TM weighings averaged over this range. The
effect of any nonlinearity essentially cancels due to the
averaging process. The accuracy of the nonlinearity correction is
described in Sec.~\ref{sec310}.

The 256 load steps were accomplished using 15 AM's with a mass of
approximately 783~\mug\ called AM1's and 15 AM's with 16 times
this mass (12,528~\mug) called AM16's. They were made from short
pieces of stainless steel wire with diameters of 0.1 mm and
0.3~mm. The wires were bent through about $70^\circ$ on both ends
leaving a straight middle section of about 6~mm. The mass of the
AM's were electrochemically etched to obtain as closely as
possible the desired masses. The RMS deviation was 1.5~\mug\ for
the AM1's and 2.3~\mug\  for the AM16's.

By weighing a TM together with various AM combinations, one
obtains the value of the TM weight simultaneously with the
linearity information. The only additional time required for this
procedure over that of weighing only the TM's is the time
necessary to change an AM combination (10 to 30~s).

\subsection{Mass Handler}
\label{sec35} 
\begin{figure}
\includegraphics[width=8cm]{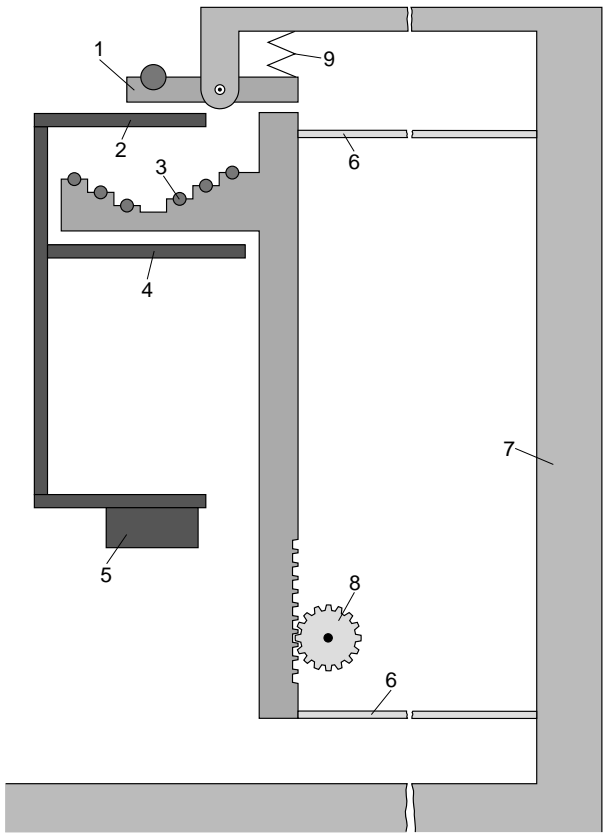}
\caption{Simplified drawing of the mass handler illustrating the
principle of operation. Legend: 1=pivoted lever pair holding a CM,
2=narrow strip to receive the CM, 3=double stair case pair holding
AM's, 4=narrow strip to receive AM's, 5=balance pan, 6 flat
spring, 7=frame, 8=stepmotor driven cogwheel, and 9=coil spring.
The pivoted-lever pair and the double-staircase pair are spaced
such that they can pass on either side of the narrow strips 2 and
4 fastened to the balance pan. The two flat springs 6 form two
sides of a parallelogram which assures vertical motion of the
double stair case pair.} \label{fig:masshandler}
\end{figure}

The mass handler is the device which placed the AM's
and the CM's on the balance or removed them from the balance. The
mass handler was designed by the firm Metrotec AG. The operation
of this device is illustrated in the somewhat simplified drawing
of Fig.~\ref{fig:masshandler} showing how the AM1's and the CM1
are placed on the metal strip attached to the balance pan. Only 6
of the 14 steps are shown in this illustration for clarity. The
portion of the handler used for the AM'16 and the CM2 (not shown)
is similar except that the AM'16 are placed on a metal strip
located below the one used for the AM1's. All of the AM1's
pictured in Fig.~\ref{fig:masshandler} are lying on the steps of a
pair of parallel double staircases. The staircases are separated
by 6~mm which is the width of the AM's between the bent regions on
both ends. The spacing between the staircases is such that they
could pass on either side of the horizontal metal strip fastened
to the balance pan as the staircases were moved up or down. The
motion of each staircase pair was constrained to the vertical
direction by a parallelogram (similar to those of the balance)
fastened to the frame of the mass handler. The staircases for
AM1's and AM16's were moved by two separate step motors located
outside the vacuum system. The step motors were surrounded by mu
metal shielding to reduce the magnetic field in the neighborhood
of the balance. Moving the staircases down deposited one AM after
another onto the metal strip. Moving the staircases up removed the
AM's lying on the strip. The steps of the staircase had hand
filed, saddle shaped indentations to facilitate the positioning of
the AM's. The heights of the steps were 2~mm and the steps on the
left side of the double staircase were displaced in height by 1~mm
from those on the right. Thus, the AM1's were alternately placed
on the balance to the left and to the right of the center of the
main pan in order to minimize the torque which they produced on
the balance.

Raising the staircase structure above the position shown in the
figure caused a rod to push against a pivoted lever holding CM1.
With this operation, CM1 was placed on the upper strip attached to
the balance. Reversing the operation allowed the spring to move
the lever in the opposite direction and remove CM1 from the
balance.

Due to the very small mass of the AM1's, difficulty was
occasionally  experienced with the AM1's sticking to one side of
the staircase or the other. The staircases were made of aluminum
and were coated with a conductive layer of tungsten carbide to
reduce the sticking probability. Sticking nevertheless did occur.
The sticking would cause an AM1 to rest partly on the staircase
and partly on the pan, thus giving a false balance reading. In
extreme cases, the AM1 would fall from the holder and therefore be
lost for the rest of the measurement. No problem was experienced
with the heavier AM16's and the CM's.

\subsection{Weighing Schedule}
\label{sec36} The experiment was planned so that the zero-point
(ZP) drift and the linearity of the balance could be determined
while weighing a TM. In principle one needs just 4 weighings
(upper and lower TM with FM's together and apart) to determine the
signal for each AM placed on the balance. Repeating these 4
weighings allows one to determine how much the zero point has
changed and thereby correct for the drift. Since there are 256 AM
values required to correct the nonlinearity of the balance, a
minimum of 2048 weighings is needed for a complete determination
of the signal corrected for ZP drift and linearity. One also
wishes to make a number of calibration measurements during the
series of measurements.

The order in which the measurements are performed influences
greatly the ZP drift correction of measurements. Changing AM's
requires only 6 to 30~s, while exchanging TM on the balance takes
about 230 s and moving the FM from one position to the other
requires about 600 s. These times are to be compared with the 80 s
required for a weighing and about 1 hr for a complete calibration
measurement (see Sec.~\ref{sec39}). One therefore wishes to
measure a number of AM values before exchanging TM, and repeat
these measurements for the other TM before changing the FM
positions or making a calibration.

The schedule of weighing adopted is based on several basic series
for the weighing of the different TM's with different FM
positions. The series are defined as follows:
\begin{enumerate}
\item An S4 series is defined as the weighing of four successive
AM values with a particular TM and with all weighing made for the
same FM positions. \item An S12 series involves three S4 series
all with the same four AM values and the same FM positions. The S4
series are measured first for one TM, then the other TM and
finally with the original TM. \item An S96 is eight S12 series,
all made with the same FM positions and with the AM values
incremented by four units between each S12 series. A TM exchange
is also made between each S12 series. An S96 series represents the
weighings with 32 successive AM values for both TM all with the
same FM positions. \item An S288 series is three S96 series, first
with one FM position, then the other and finally with the original
FM position. A calibration measurement is made at the beginning of
each S288 series. Thus, the S288 series represents the weighings
with 32 successive AM values for both TM's and both FM positions
and includes its own calibration. \item An S2304 series is made up
of eight S288 series with the AM values incremented by 32 between
each S288 series. An S2304 series completes the full 256 AM values
with weighings of both TM's and both FM positions.
\end{enumerate}

A total of eight valid S2304 series was made over a period of 43
days. Alternate S2304 series were intended to be made with
increasing and decreasing AM values. Unfortunately, the restart
after a malfunction of the temperature stabilization in the
measuring room was made with the wrong incrementing sign. This
resulted in five S2304 series being made with increasing AM values
and three with decreasing.

\subsection{Analysis of the Weighings}
\label{sec37} 
\begin{figure}
\includegraphics[width=8.5cm]{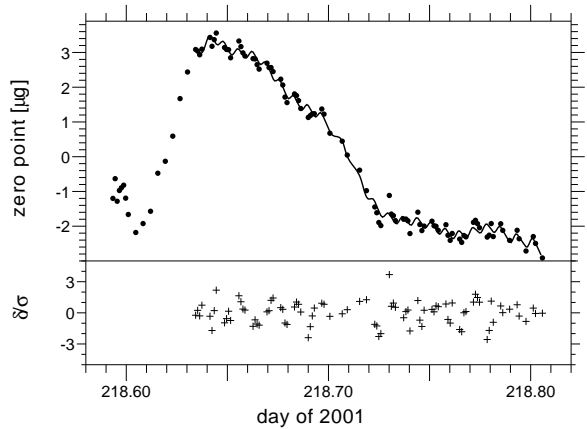}
\caption{The zero-point variation as a function of time for a
typical S96 series including calibration is shown in the upper
part of this figure. The solid curve is the fit function starting
after the last dummy weighing. The fit function for this S96
series has 76 degrees of freedom. The normalized residuals
$\delta/\sigma$ are shown in the lower plot. The normalization of
the residuals has been chosen such that their rms value is 1.}
\label{fig:zeropoint}
\end{figure}

In ref.~\cite{Sa02,Sb02}, the so called ABA method
was used to analyze the data obtained from the balance and thereby
obtain the difference between the mass of the A and B TM. This
method assumes a linear time dependence of the weight that would
be obtained for the A TM at the time when the B TM was measured
based on the weights measured for A at an earlier and a later
time. However, a careful examination of the data showed that the
curvature of the ZP drift was quite large and was influenced by
the previous load history of the balance. This indicated that the
linear approximation was not a particularly good approximation. We
have therefore reanalyzed the data using a fitting procedure to
determine a continuous ZP function of time for each S96 series.
The data and fit function for a typical S96 series starting with a
calibration measurement is shown in Fig.~\ref{fig:zeropoint}. The
procedure used to determine the ZP data and fit are described in
the following. The criterion for a valid weighing is described in
Sec.~\ref{sec38}.

The data of Fig.~\ref{fig:zeropoint} show a slow rise during the
first hour after the calibration measurement followed by a
continuous decrease with a time constant of several hours. These
slow variations are attributed to thermal variations resulting
principally from the different loading of the balance during the
calibration measurement. Superposed on the slow variations are
rapid variations which are synchronous with the exchange of the
TM's. The rapid variations peak immediately after the TM exchange
and decrease thereafter with a typical slope of 0.3 ~\mug\ /hr.
The cause of the rapid variations is unknown.

The data employed in the ZP determination were the weighings of
the upper and lower TM's for the S96 series. The known AM load for
each weighing was first subtracted to obtain a net weight for
either TM plus the unknown zero-point function at the time of each
weighing.  A series of Legendre polynomials was used to describe
the slow variations of the zero-point function. A separate $P_0$
coefficient was employed for each TM. The rapid variations were
described by a sawtooth function starting at the time of each TM
exchange. The fit parameters were the coefficient of the Legendre
polynomials and the amplitude of the sawtooth function. The
sawtooth amplitude was assumed to be the same for all rapid peaks
of an S96 series. The sawtooth function was used principally to
reduce the $\chi^2$ of the fit and had almost no effect on the
results obtained when using the ZP function. All parameters are
linear parameters so that no iteration is required. The actual ZP
function is the sawtooth function and the polynomial series
exclusive of the time independent terms (i.e. the sum of the
coefficients times $P_n(0)$ for even $n$).

Such calculations were made for various numbers of Legendre
coefficients in the ZP function. It was found that the
gravitational signal was essentially constant for a maximum order
of Legendre polynomials between 8 and 36. In this range of
polynomials, the minimum calculated signal was 784.8976(91)~\mug\
for a maximum order equal to 22 and a maximum signal of
784.9025(93)~\mug\ for a maximum order equal to 36 (i.e. a very
small difference). In all following results, we shall use the
signal 784.8994(91)~\mug\ obtained with a maximum polynomial order
of 15.

It has been implicitly assumed in the above ZP determination that
the AM load values were known with much better accuracy than the
reproducibility of the balance producing the data used in the ZP
fit. Although the AM values were sufficiently accurate for
determining the general shape of the ZP function, their relative
uncertainties were comparable to the uncertainties of the balance
data used in the fit. The $P_0$ mass parameters of the TM's
obtained from the fit were therefore not used for the TM
differences at the two FM positions which are needed in order to
determine the gravitational signal. Instead, the value of the ZP
function was subtracted from each weighing, and an ABA mass
difference was determined for each triplet of weighings having the
same AM load. Since the mass of the AM's do not occur in this TM
difference, they do not influence the calculation. The ABA
calculation is valid for this purpose since the ZP corrected
weighings have essentially no curvature. Such TM differences
determined for the apart-together positions of the FM's are then
used to calculate the gravitational signal.

\begin{figure}
\includegraphics[width=8.5cm]{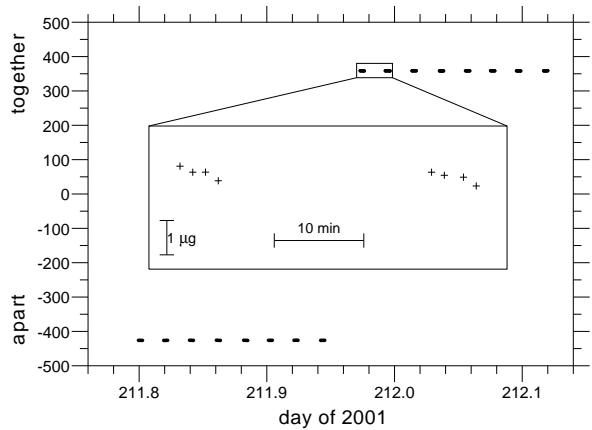}
\caption{The measured weight difference in $\mu g$ between TM's
obtained from an S288 series. The magnified insert shows the
individual TM differences which are not resolved in the main part
of the figure.} \label{fig:S288}
\end{figure}

The TM differences for the apart-together positions of the FM's as
a function of time are shown for a ZP corrected S288 series in
Fig.~\ref{fig:S288}. Individual data points are resolved in the
magnified insert of this figure. Each data point is the B member
of a TM difference obtained from an ABA triplet in which all
weighings have the same load value.

\begin{figure}
\includegraphics[width=8.5cm]{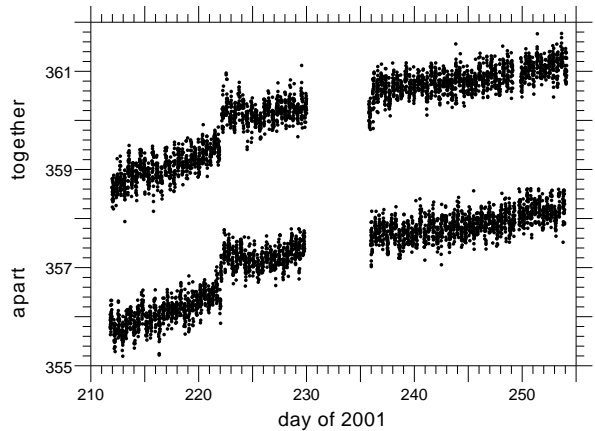}
\caption{The measured weight difference in $\mu g$ between TM's
for the FM's positions apart and together. The measured values for
the FM apart have been displaced 782 $\mu g$ in order to show both
types of data in the same figure. } \label{fig:tmdiff}
\end{figure}

All of the TM differences (ZP corrected) for the entire experiment
are shown in Fig.~\ref{fig:tmdiff}. The data labelled apart have
been shifted by 782~\mug\ in order to allow both data sets to be
presented in the same figure. A slow variation of 2.5~\mug\ in
both TM differences occurred during the 43 day measurement. Also
seen in this figure is a 0.7~\mug\ jump which occurred in the data
for both the apart and together positions of the FM's on day 222.
The slow variation is probably due to sorption-effect differences
of the upper and lower TM's. The jump was caused by the loss or
gain of a small particle such as a dust particle by one of the
TM's. In order to determine the gravitational signal, an ABA
difference was calculated for apart-together values having the
same AM load. The slow variation seen in Fig.~\ref{fig:tmdiff} is
sufficiently linear so that essentially no error results from the
use of the ABA method. The jump in the apart-together differences
caused no variation of the gravitational signal.

\begin{figure}
\includegraphics[width=8.5cm]{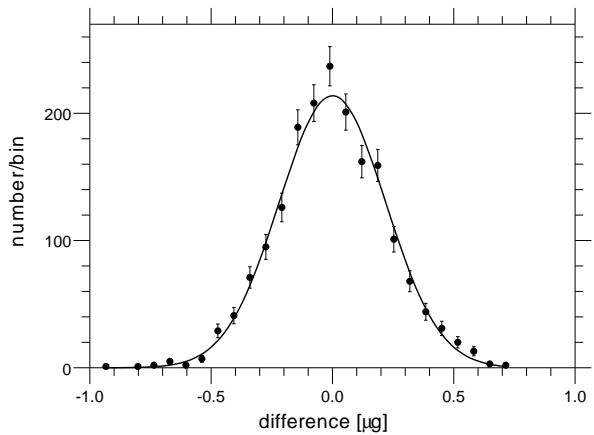}
\caption{Binned data for the FM apart-together weight differences
(points) and a fitted Gaussian function (curve) shown as a
deviation from the mean difference. Poisson statistics were used
to determine the uncertainties.} \label{fig:histo}
\end{figure}

In Fig.~\ref{fig:histo} is shown a plot of the binned difference
between the FM apart-together positions for all valid data (see
Sec.~\ref{sec38}. The differences were determined using the ABA
method applied to weighings made with the same AM loads. Also
shown in the figure is a Gaussian function fit to the data. The
data are seen to agree well with the Gaussian shape which is a
good test for the quality of experimental data. The
root-mean-square (RMS) width of the data is 1.03 time the width of
the Gaussian function. The true resolution for these weighings may
be somewhat different than shown in Fig.~\ref{fig:histo} due to
the fact that the data have not been corrected for nonlinearity of
the balance (see Sec.~\ref{sec310}) and for correlations due to
the common ZP function . Nevertheless, these effects would not be
expected to influence the general Gaussian form of the
distribution.

\begin{figure}
\includegraphics[width=8.5cm]{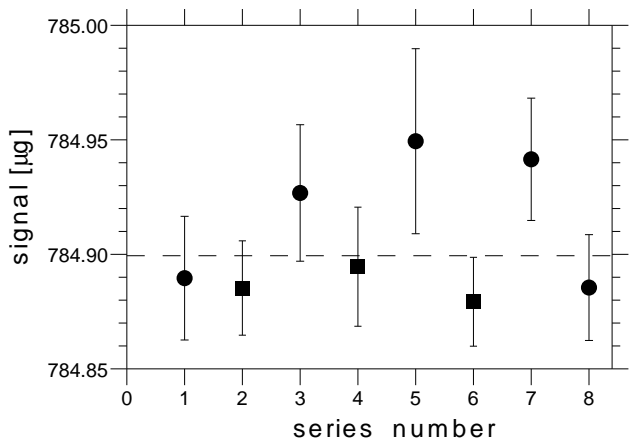}
\caption{The average signal for each of the eight S2304 series.
Series with increasing load are shown as circles. Series with
decreasing load are shown as squares. The dashed line is the
average of all eight series.} \label{fig:cycles}
\end{figure}

A plot of the signal obtained for the S2304 series with increasing
and decreasing load is shown in Fig.~\ref{fig:cycles}. The average
signal for increasing load is 784.9121(125)~\mug\ and the average
for decreasing load is 784.8850(133)~\mug. The common average for
both is 784.8994(91)~\mug. The averages for increasing load and
for decreasing load lie within the uncertainty of the combined
average. This shows that the direction of load incrementing did
not appreciably influence the result.

Although the weighings making up an S96 series are correlated due
to the common ZP function determined for each S96 series, the
results of each S96 series, in particular the TM parameter, are
independent. The 32 signal values obtained from the three S96
series making each S288 series are also independent. However,
since the nonlinearity correction (see Sec.~\ref{sec310}) being
employed is applicable only to an entire S2304 series (not to
individual S288 series), it is only the eight S2304 series which
should be compared with one another. This restricts the way in
which the average signal is to be calculated for the entire
measurement, namely the way in which the data are to be weighted.

We have investigated two weighting procedures. In the first, each
S2304 series average was weighted by the number of valid triplets
in that series. This assumes that the weighings measured in all
S2304 series have the same a priori accuracy. In the second
method, it was assumed that the accuracy for each weighing in a
series was the same but might be different for different series.
We believe the second method is the better method since it takes
into account changes that occur during the long, 43 day
measurement (e.g. the not completely compensated effects of
vibration, tidal forces and temperature). The averages obtained
with the two methods differ by approximately 6~ng with the second
method giving the smaller average signal. This is a rather large
effect. It is only slightly smaller than the statistical
uncertainty of 9 to 10~ng obtained for either method. In the rest
of this work, we shall discus only the results obtained with the
second method.

\subsection{Criterion for Valid Data}
\label{sec38}
Two tests were used to determine whether a measured weighing was
valid. An on-line test checked whether the $\chi^2$ value of the
fit to the pendulum oscillations was reasonable. A large value
caused a repeat of the weighing. After two repeats with large
$\chi^2$, the measurement for this AM value was aborted. An
aborted weighing usually indicated that the AM was resting on the
mass handler and on the balance pan in an unstable way.

A more frequent occurrence was that of an AM which rested on both
the mass handler and the balance was almost stable thereby giving
a reasonable $\chi^2$. In order to reject such weighings, an
off-line calculation was made to check whether the measured weight
was within 10~\mug\ of the expected weight. The statistical noise
of a valid weighing was typically about 0.15~\mug\ (see
Fig.~\ref{fig:histo} showing ABA difference involving 3
weighings). Excursions of more than 10~\mug\ were thus a clear
indication of a malfunction.

This off-line test is somewhat more restrictive than the off-line
test employed in our original analysis. In the original analysis,
a check was made only to see that the weight difference between
the TM's for equal AM loadings was reasonable. The more
restrictive test used in the present analysis resulted in the
rejection of the S2304 series at the time when the room
temperature stabilizer was just beginning to fail. It was also the
reason for not including the tantalum TM-measurements in the
present analysis. In the eight S2304 series accepted for the
determination of the gravitational constant, approximately 8\% of
the expected zero-point values could not be determined due to at
least one of the three weighings at each load value being rejected
by the test for valid weighings.

\subsection{Calibration Measurements}
\label{sec39} A coarse calibration of the balance  was made
periodically during the gravitational measurement (before each
S288 series) using two calibration masses each with a weight of
approximately 100~mg. A correction to the coarse calibration
constant due to the nonlinearity of the balance will be discussed
in Sec.~\ref{sec310}. The two CM's used for the coarse calibration
were short sections of stainless steel wire. The diameter of CM1
was 0.50(1)~mm and that of CM2 was 0.96(1)~mm. The surface area of
CM1 was approximately 1~cm$^2$ and that of CM2 was 0.5~cm$^2$. The
CM's were electrochemically etched to the desired mass and then
cleaned in an ethanol ultrasonic bath. The mass of each CM was
determined at METAS (Metrology and Accreditation Switzerland) in
air with an absolute accuracy of 0.4~\mug\  or a relative
uncertainty of 4~ppm. The absolute determinations of the CM masses
were made before and after the gravitational measurement with
copper TM's and after the second measurement with tantalum TM's.
Only the first measurement was used to evaluate the coarse
calibration constant employed in the measurement with copper TM's.
As will be discussed below, the second and third measurement were
used for the measurement with copper TM's only to check the
stability of the CM's.

A calibration measurement involved either TM and one of following
seven additional loads: (1) CM1 alone, (2) CM2 alone, (3) again
CM1 alone, (4) empty balance, (5) CM1+CM2, (6) empty balance and
(7) CM1+CM2 and nine so called dummy weighings. These measurements
were made with no AM's on the balance. After the seventh weighing,
a series of nine dummy weighings alternating between upper and
lower TM's were made with the AM load set to the value for the
next TM weighing. The dummy weighings were made in order to allow
the balance to recover from the large load variations experienced
during the calibration measurement and thereby come to an
approximate equilibrium value before the next TM weighing.
Calibrations were made alternately with the upper and lower TM's
as load. Calibration measurements were made about twice a day.
Including the dummy weighings, each calibration required about 50
min.

A three-parameter least squares fit was made to the calibration
weighings labelled 4,5,6 and 7 above. The fit thereby determined
best values for the balance ZP, the slope of the ZP and a
parameter representing the effective ZP corrected reading of the
balance for the load CM1+CM2. This third parameter is of
particular interest since the coarse calibration constant is
determined from the known mass of CM1+CM2 (measured by METAS)
divided by this parameter. Therefore, the results of the
least-squares fit to each set of calibration data gave a value for
the coarse calibration constant which then was used to convert the
balance output of the S288 series to approximate mass values. An
ABA analysis of the first three weighings of each set of
calibration data was also made in order to determine the
difference in mass between CM1 and CM2.

The absolute masses obtained for CM1 and CM2 as determined by
METAS are given in columns 2 and 3 of Table~\ref{tab:cm}. Also
shown in Table~\ref{tab:cm} (column 4) are the mass difference
between CM1 and CM2 as obtained from the METAS  measurement in air
and the average of our CM measurement in vacuum. The mass
differences between CM1 and CM2 measured in vacuum are
particularly useful in checking for any mass variation of the
CM's.

\begin{table}[h]
\caption{The mass of the CM's as measured by METAS and the CM1-CM2
mass differences measured in air at METAS and in vacuum during the
gravitational measurements at PSI. All values are given in $\mu
g$.}
\label{tab:cm}
\begin{ruledtabular}
\begin{tabular}{l l l l }
\multicolumn{1}{c}{Date} & \multicolumn{1}{c}{CM1} &
\multicolumn{1}{c}{CM2} &
\multicolumn{1}{c}{Difference} \\
\hline
Feb 6, 01 & 100,270.30(40) & 100,263.90(40) & 6.40(60)\\
Jul. - Sep., 01  & \multicolumn{2}{c}{in vacuum} & {5.853(19)}\\

Nov. 29, 01 & 100,270.20(35) &100,262.90(35) & 7.30(50)\\
{Jan. - Mar., 02}  & \multicolumn{2}{c}{in vacuum} & {7.269(29)}\\

{Apr. - May, 02}  & \multicolumn{2}{c}{in vacuum}  & {7.496(25)}\\
May  27, 02        & 100,270.01(35)&100,262.97(35) &7.04(50)\\
\end{tabular}
\end{ruledtabular}
\end{table}

It is seen that CM2 mass decreased by 1.00(53)~\mug\ between the
first and second METAS measurements while the mass of CM1 was
essentially the same in all three measurements. From the mass
difference values in air and vacuum it is clear that the change
occurred after the measurements with copper TM's ended in
Sept.~2001 and before the weighing at METAS in Nov.~2001 which
preceded  the start of the tantalum measurements. We ascribe this
change of CM2 to either the loss of a dust particle or perhaps a
piece of the wire itself. The loss of a piece of the wire was
possible since the wire used for the CM's had been cut with a wire
cutter and there could have been a small broken piece that was not
bound tightly to the wire. For this reason only the values given
for the first weighing of the CM's were used to determine the
coarse calibration constant used for the measurement with copper
TM's.

\begin{figure}[b]
\includegraphics[width=8.5cm]{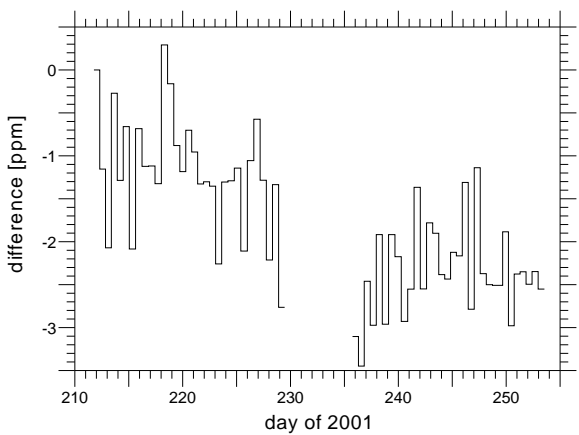}
\caption{The change of the effective balance reading for the load
CM1+CM2 as a function of time relative to its value on the first
day. No valid measurements were made between day 229 and 235.}
\label{fig:cm}
\end{figure}

A plot of the relative change of the effective ZP corrected
balance reading corresponding to the load CM1+CM2 is shown in
Fig.~\ref{fig:cm}. It is seen that it changed by only a few ppm
over the 43 days of the measurement. A linear fit made to these
data results in a slope equal to -0.044(6)~ppm/day which is
equivalent to a mass rate variation of -0.0088(12)~\mug\;cm$^{-2}$
\;d$^{-1}$. The uncertainty was obtained by normalizing $\chi^2$
of the fit to the degrees of freedom (DOF).

The slow variation of the effective balance reading for the load
CM1+CM2 seen in Fig.~\ref{fig:cm} could be due either to a change
of the balance sensitivity, to a decrease in the mass of CM1+CM2
due to the removal of a contaminant layer from the CM's in vacuum
or to a combination of both causes. A variation of the balance
sensitivity would have essentially no effect on the analysis of
the weighing for the gravitational measurement as the coarse
calibration constant used for the analysis was determined from the
balance parameter for each S288 series. However, a variation of
the mass of CM1+CM2 would result in an error in the analysis since
the mass would not be the value measured by METAS shown in
Table~\ref{tab:cm}.

\begin{figure}
\includegraphics[width=8.5cm]{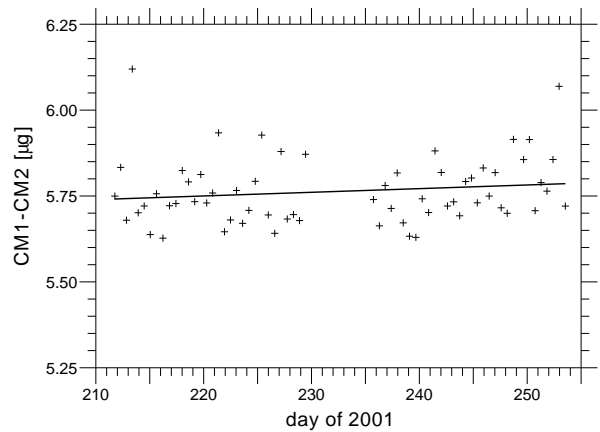}
\caption{The mass difference of the CM's  as a function of time
and the linear fit function.} \label{fig:cmdiff}
\end{figure}

In order to investigate this problem, we have examined the
difference between the balance readings for CM1 and CM2. This
difference is proportional to the surface areas of CM1 and CM2
which differ by approximately a factor of 2 (CM1 area=1~cm$^2$ and
CM2 area=0.5~cm$^2$). The balance reading difference is only
slightly dependent upon the coarse calibration constant so that it
represents essentially the mass difference itself. In
Fig~\ref{fig:cmdiff} is shown the measured mass difference as a
function of time  during the gravitational measurement. Also shown
is a linear function fit to these data. The slope parameter of the
fit results in a rate of increase per area equal to
0.0021(18)~\mug\;cm$^{-2}$ \;d$^{-1}$. The uncertainty has been
determined by normalizing $\chi^2$ to the DOF. The sign of the
slope is such that the CM with the larger area has the larger rate
of increase. A mass difference variation (CM1-CM2) would require a
slope of -0.0088(12)~\mug\;cm$^{-2}$ \;d$^{-1}$. The measured
slope of the effective balance reading for the load CM1+CM2
clearly excludes such a large negative slope as assumed for a mass
variation. We therefore conclude that the variation of this
parameter is due primarily to the sensitivity variation of the
balance.

We note that Schwartz \cite{Sch94b} has also found a mass increase
for stainless steel samples in a vacuum system involving a rotary
pump, a turbomolecular pump and a liquid nitrogen cold trap. His
samples were 1~kg masses with surface areas differing by a factor
of 1.8. He measured the thickness of a contaminant layer using
ellipsometry as well as the increase in weight of the sample
during pumping periods of 1.2~d and 0.36~d. The rate of mass
increase per area which he reports is approximately a factor of 5
larger than the value we find. No explanation for this difference
can be made without a detailed knowledge of the partial pressures
of the various contaminant gases in the two systems and the
surface properties of the samples employed.

There still remains the possibility that a rapid removal of an
adsorbed layer such as water might have occurred between the
absolute determination of the CM masses in air at METAS and the
gravitational measurement in vacuum (i.e. during the pump down of
the system). Schwartz \cite{Sch94a} has measured the mass
variation per unit area of 1~kg stainless steel objects in air
with relative humidity between 3\% and 77\%. He \cite{Sch94b} also
has measured the additional mass variation per area  due to
pumping the system from atmospheric pressure at 3\% relative
humidity down to $5\times 10^{-3}$~Pa. His samples were first
cleaned by wiping them with a linen cloth soaked in ethanol and
diethylether and then ultrasonic cleaning in ethanol. After
cleaning, they were dried in a vacuum oven at 50~$^\circ$C. For
these cleaned samples, the weight change found for 3\% to 50\%
humidity variation was 11.5~ng \;cm$^{-2}$ with an additional
change of 29~ng \;cm$^{-2}$ in going from 3\% relative humidity in
air to vacuum (total change of 40.5~ng cm$^{-2}$). Similar
measurements with "uncleaned" samples gave a total change of
80~ng\;cm$^{-2}$. The variation due to the cleanliness of the
samples was much larger than the difference found for the two
types of stainless steel investigated and the effect of improving
the surface smoothness (average peak-to-valley height equal to
0.1~\mum\  and 0.024~\mum). Since the cleaning procedure used for
our CM's and their smoothness were different than the samples used
by Schwartz, we have employed the average of Schwartz's "cleaned"
and "uncleaned" objects for estimating the mass change of our
CM's. Based on these data, the relative mass difference found for
both CM's together as measured in air having 50\% humidity and in
vacuum was 0.5~ppm. We assign a relative systematic uncertainty of
this correction equal to the correction itself.

\subsection{Nonlinearity Correction}
\label{sec310}
By nonlinearity of the balance, one is referring to the variation
of the balance response function with load, that is, the degree to
which the balance output is not a linear function of the load. The
nonlinearity of a mass comparator similar to the one employed in
the present work has been investigated \cite{Re00} by the firm
Mettler-Toledo. It was found that besides nonlinearity effects in
10~g load intervals, there was also a fine structure of the
nonlinearity in the 0.1~mg load interval which would be important
for the accuracy of the present measurement. It is the
nonlinearity of our mass comparator in the particular load
interval less than 0.2~g involved in the present experiment that
we wish to determine.

One expects the nonlinearity of the balance used in this
experiment to be small; however, it should be realized that a 200
mg test mass (two 100~mg CM's) required for having an accurately
known test mass for calibration purposes is over 250 times the
size of the gravitational signal that one wishes to determine. In
addition, the statistical accuracy of the measured gravitational
signal is some 30 times better than the specified accuracy
(2~\mug) of the unmodified commercial balance. One therefore has
no reason to expect the nonlinearity of the balance to be
negligible with this precision. In Sec.~\ref{sec31} we have
presented the general idea that the measurements with 256 AM
values tends to average out the effect of any nonlinearity. We
wish now to give a more detailed analysis of this problem.

\begin{figure}
\includegraphics[width=8.5cm]{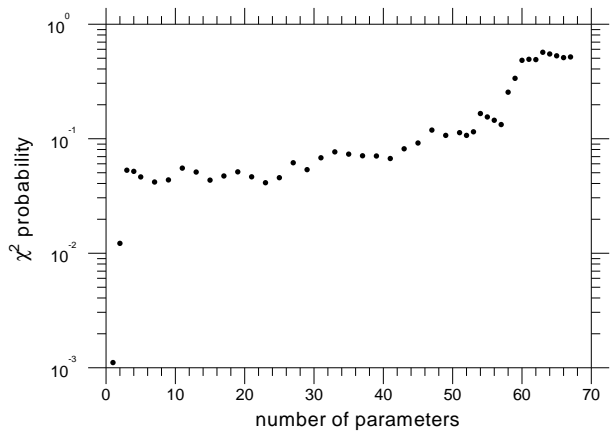}
\caption{The $\chi^2$ probability as a function of the number of
parameters.} \label{fig:chivsl}
\end{figure}

The correction for nonlinearity makes use of an arbitrary response
as a function of the load. Since the two TM's are essentially
equal ($<400$~\mug\  difference), the variation of the response
function can be thought of as being a function of the additional
load due to the AM's. Although a power series or any polynomial
series would suffice for this function, we have for convenience
used a series of Legendre polynomials
\begin{displaymath}
  f(u)=\sum_{\ell=0}^{Lmax} a_\ell P_{\ell} (2 u/u_{max}-1).
\end{displaymath}
The coefficients of $P_\ell$ are chosen subject to the two
conditions that (1) $f(u)=0$ for no load and (2) $f(u)=C$ for
$u=C$ where $C$ is the weight of the two CM's together. These two
conditions represent the sensitivity of the balance over the 0.2~g
range of the calibration (i.e. the coarse calibration). The value
of the maximum load $u_{max}$ in the present measurements was very
nearly $C$. Substituting the above conditions into the response
function, one obtains for the lowest two coefficients the
expressions
\begin{displaymath}
  a_0=C/2-\sum_{even \; \ell=2}^{Lmax} a_\ell
\end{displaymath}
and
\begin{displaymath}
  a_1=C/2-\sum_{odd \; \ell=3}^{Lmax} a_\ell.
\end{displaymath}

One can then minimize
\begin{displaymath}
   \chi^2=\sum_{n=1}^N \left [f(u_n+s)
   -f(u_n)-y_n\right]^2 \sigma_n^{-2}
\end{displaymath}
and thereby determine best values for the parameters $s$ and
$a_\ell$ for $\ell=1$ to $Lmax$. The $y_n$ are the measured
balance signal for the load values $u_n$, $s$ is the load
independent signal and $N$ is the number of different loads with
valid measurements. The error $\sigma_n$ for the load value $u_n$
is the load-independent intrinsic noise of the balance $\sigma_0$
for a single weighing divided by the square root of the number of
weighings for the load $u_n$. The value of $Lmax$ must be chosen
large enough to describe the response function accurately. All of
the parameters in the fit are linear parameters with the exception
of $s$. Thus, there is no difficulty in extending the fit to a
large number of parameters since only the nonlinear parameter must
be determined by a search method.

In order to determine $Lmax$, we calculate the $\chi^2$
probability \cite{Ba89} (often referred to as confidence level) as
a function of $Lmax$. This requires an approximate value for the
intrinsic noise of the balance $\sigma_0$. The value of $\sigma_0$
sets the scale of the $\chi^2$ probability but does not change the
general shape of the function. One can obtain a reasonable
approximation for $\sigma_0$ by setting $\chi^2$ equal to the DOF
obtained for a large number of parameter. We have arbitrarily  set
$\chi^2$ equal to the DOF for 61 parameters. The $\chi^2$
probability as a function of the maximum number of parameters is
shown in Fig.~\ref{fig:chivsl}. It is seen that the $\chi^2$
probability reaches a plateau near this maximum number of
parameters.

Starting from a low value of $10^{-4}$ for one parameter, the
$\chi^2$ probability rises rapidly to a value of 0.05 for three
parameters. It remains approximately constant at this value up to
57 parameters where it rises sharply to reach a plateau of
approximately 0.5 at 60 parameters and above. The fit parameter
representing the signal corrected for nonlinearity of the balance
was essentially constant over the entire range of parameters with
a variation of less than $\pm 1.3$~ng. The signal for one
parameter representing complete linearity was 784.8994~\mug. The
signal of the plateau region from 60 to 67 parameters was
784.9005~\mug\ with a statistical uncertainty of 5.5~ng. In this
region the signal varied by less than 0.2~ng. We therefore take
the nonlinearity correction of the measured signal to be
1.1(5.5)~ng (i.e. the difference between the signal using one
parameter as would be obtained with no correction and the average
value obtained for 60 to 67 parameters).

\begin{figure}
\includegraphics[width=8.5cm]{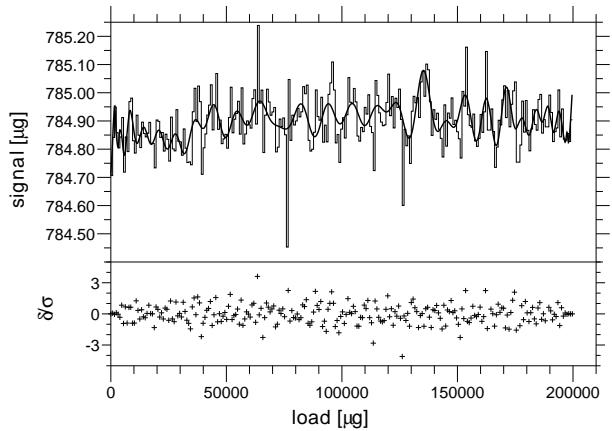}
\caption{Signal and fit function employing 60 parameters as a
function of load. The data are shown as a stepped line. The fit is
the smooth curve. The lower plot shows the normalized residuals.
Residuals were divided by the relative uncertainty of each point.
The normalization has been chosen such that the rms value of the
residuals is 1.} \label{fig:sigfit}
\end{figure}

The nonlinear signal and fit as a function of load determined for
60 parameter is shown in Fig.~\ref{fig:sigfit}. The function shows
many narrow peaks with widths of 3 to 10 load steps and with
amplitudes of roughly 0.1~\mug.  In principle one could use this
response function to correct the individual weighings with various
loads; however, we prefer to use the signal as corrected for
nonlinearity over the entire range of measurements as described
above. The variation of the response function indicates that a
measurement made at an arbitrary load  value could be in error by
as much as $\pm 130$~ng assuming the response to be linear. This
is to be compared with the assumed uncertainty in ref.~\cite{Sb98}
due to nonlinearity of $\pm 200$~ng.

\subsection{Correction of the TM-Sorption Effect}
\label{sec311}
Moving the FM's changed slightly the temperature of the vacuum
tube surrounding the TM's. These temperature variations were due
to changes in the air circulation in the region of the vacuum tube
as obstructed by the FM's. An increase of the wall temperature of
the tube caused adsorbed gases to be released which were then
condensed onto the TM. Since the temperature variation was
different in the regions near the upper and lower TM's, this
resulted in a variation of the weight difference between the upper
and lower TM's (i.e. a "false" gravitational signal).

\begin{figure}
\includegraphics[width=8.5cm]{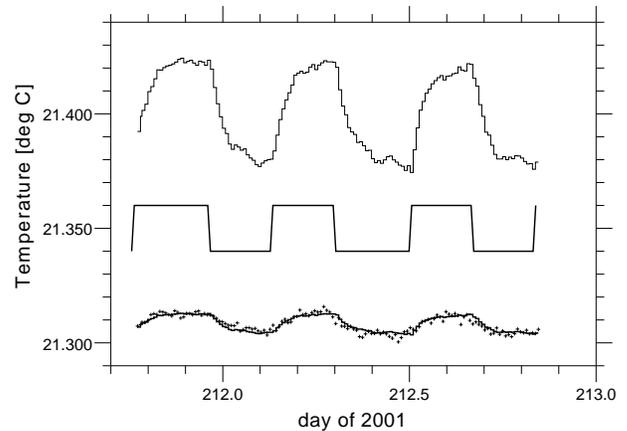}
\caption{Temperatures of the vacuum tube measured at the position
of the TM's. The upper curve is the temperature at the position of
the upper TM. The square wave in the middle section of the plot
indicates the FM motion. The data (crosses) for the lower TM and
fit function (solid line) are shown in the lowest section of the
figure.} \label{fig:temp}
\end{figure}

The temperature variation at the positions of the upper and lower
TM's during one day of the gravitational measurement is shown in
Fig.~\ref{fig:temp} along with a curve representing the FM motion.
The peak-to-peak temperature variation was approximately
$0.04^\circ \mbox{C}$ at the upper position and $0.01^\circ
\mbox{C}$ at the lower position. The shape of the temperature
variation at the upper position was used as a fit function
(employing an offset and an amplitude parameter) to obtain a
better determination of the temperature variation at the lower
position. There were 32 one-day measurements of the temperature
variations during the gravitational measurement. The average
amplitude at the lower position determined from these 32
measurements was $0.0138(2)^\circ \mbox{C}$.

\begin{figure}
\includegraphics[width=8.5cm]{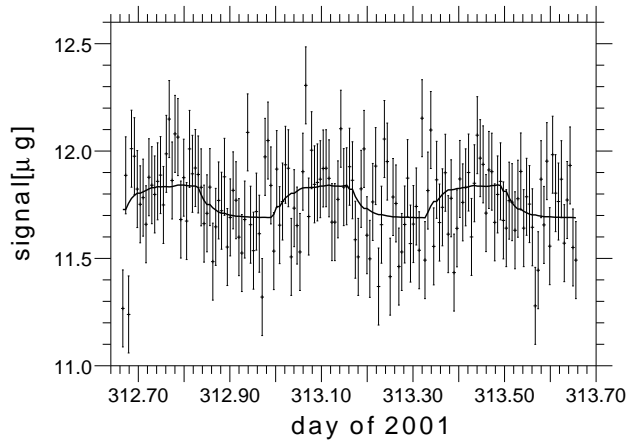}
\caption{Weight difference between TM's as a function of time for
a temperature variation roughly 10 times that of the gravitational
measurement. The solid curve is the best fit of the temperature
variation difference at upper and lower TM positions. For the
purpose of this plot, an arbitrary offset of the weight difference
between upper and lower TM has been employed.}
\label{fig:sorption}
\end{figure}

The signal produced by these temperature variations was small and
therefore not directly measurable with the balance in a reasonable
length of time. The procedure that was employed to determine this
temperature dependent signal was to use four electrical heater
bands to produce a variation of the temperature distribution along
the vacuum tube that was a factor of approximately seven larger
than the variation resulting from the motion of the FM's. The
bands were positioned 30 cm above and below the positions of the
upper and lower TM's. The heater windings were bifilar to avoid
magnetic effects. The heater power (less than 3~W total) was
turned off and on with the same 8-hour period as the FM motion and
produced essentially no change in the average temperature of the
vacuum tube in the day-long measurement. The FM's were not moved
during the measurements with heaters. The signal (TM weight
difference as determined with the balance) obtained during a one
day measurement with heaters is shown in Fig.~\ref{fig:sorption}.
The shape of the fit function (employing an offset and an
amplitude parameter) shown in this figure was obtained from the
variation of the temperature difference at the upper and lower
positions of the TM's. The signal obtained from the fitting
procedure was 0.114(40)~\mug.

In order to scale the heater produced signal to that resulting
from the FM motion during the gravitational measurement, we make
the simplifying assumption that the signal variation  is
proportional to the temperature variation  at the upper TM
position minus the temperature variation  at the lower TM
position. The term variation in this statement refers to the
variation about its mean value. One uses the temperature
difference since the signal is defined as the difference between
TM weighings.

With just four heater bands it was not possible to obtain a
variation of the  temperature distribution along the vacuum tube
that was exactly a constant factor times that of the FM motion.
For the best adjustment that we were able to obtain, the ratio of
the heater produced temperature variation to the FM produced
variation was 7.1 at the upper position and 9.2 at the lower. The
ratio for the variation of the temperature difference at the upper
and lower positions relative to the gravitational values was 6.8.
These ratios are based on the peak-to-peak amplitudes obtained for
the fitted functions. The scaling factor for the temperature
difference ratio is the reciprocal of the temperature difference
ratio or 0.147. This results in a scaled signal of 0.0168(58)~\mug\
where the uncertainty is the statistical uncertainties of the
measured signal and the scaling factor. The scaled signal ("false"
signal) is to be subtracted from the total signal measured in the
gravitational experiment.

In order to check our assumption regarding the scaling factor, we
have made four additional one-day measurements in which the
temperature variations were very different from that produced by
the FM motion. The object of these measurements was to determine
whether the scaled signals obtained with the heaters were
consistent with one another when calculated with the assumed
scaling factors. The most extreme distribution involved a
temperature variation of the lower TM which was even larger
(factor of 4) than that of the upper TM. The signals obtained in
all of the test measurements were consistent with each other
within their statistical uncertainties (relative uncertainties of
approximately 30~\%). We therefore conclude that the assumption
used for scaling the signals was sufficiently accurate for the
present purpose. Nevertheless, we assign a systematic uncertainty
to the scaled signal equal to its statistical uncertainty of
5.8~ng (relative systematic uncertainty of the "false" signal is
35~\%).

\subsection{Magnetic Forces on the Test Masses}
In the absence of a permanent magnetization, the $z$ component of
force on the TM due to a magnetic field can be calculated from
\begin{displaymath}
  F_z=-\mu_0 \chi_m V H \frac{\partial H}{\partial z}
\end{displaymath}
where $V$ denotes the volume of the TM, $\chi_m$ is its magnetic
susceptibility and $H$ is the magnetic field intensity. The
magnetic properties of the TM's were measured by METAS. No
permanent magnetization was found ($<0.08$~A/m). The magnetic
susceptibility was $4\times 10^{-6}$ for the copper TM's. The
magnetic field intensity for both positions of the FM's was
measured at cm intervals along the axis of the vacuum tube at the
positions occupied by the TM's using a flux gate magnetometer. The
difference of $F_z$ for the FM positions obtained from these data
was 0.01~ng  which is a negligible correction to the measured
gravitational signal.

\subsection{Tilt Angle of Balance}
Since the weight of the TM's and the weight of the CM's both
produce forces on the balance arm in the vertical direction, a
small angle between the balance weighing direction and the
vertical produces no error in the weighing of the TM's. However,
if the balance weighing direction is correlated with the motion of
the FM's, a systematic error in the measured gravitational signal
will result. Sensitive angle monitors were mounted on the base of
the balance. No angle variation correlated with the motion of the
FM's was found with a sensitivity of 100~nrad. Since the
sensitivity of the balance varies with the cosine of the angle
(near 0 rad), this limit is completely negligible. For a balance
misalignment of 0.01~rad relative to vertical and a correlated
variation of 100~nrad with respect to this angle due to the FM
motion, the relative signal variation is approximately 0.001~ppm.

\section{Determination of the Mass-Integration Constant}

\label{sec4}
One must relate the gravitational constant to the measured
gravitation signal. This involves integrating an inverse square
force over the mass distribution of the TM's and FM's. The
gravitational force $F_z$  in the z (vertical) direction on a
single TM produced by both FM's is given by
\begin{equation}
  F_z=G  \int\int \frac {\bf{e_z} \cdot(\bf{r_2-r_1}) \; \it{dm_1
  dm_2}}{|\bf{r_2-r_1}|^{\it3}}
\end{equation}
where $\bf{e_z}$ is a unit vector in the $z$ direction, $\bf{r_1}$
and $\bf{r_2}$ are vectors from the origin to the mass elements
$dm_1$ of the TM and $dm_2$ of the FM's and $G$ is the
gravitational constant to be determined. The mass-integration
constant is the double integral in Eq. (1) multiplying $G$.
Actually, the mass-integration constant for the present experiment
is composed of four different mass-integration constants, namely
those for the upper TM and lower TM with the FM's together and
apart. We shall use as mass-integration constant the actual
constant multiplied by the 1986 CODATA value of $G$
($6.67259~\Gunit\) and give the result in dimensions of grams
"force" (i.e. the same dimensions as used for the weighings).

The objects contributing most to $F_z$ (TM's, FM tanks and the
mercury) have very nearly axial symmetry which greatly simplifies
the integration. Parts which do not have axial symmetry were
represented by single point masses for small parts and multiple
point masses for larger parts. For axial symmetric objects, we
employ the standard method of electrostatics for determining the
off-axis potential in terms of the potential and its derivatives
on axis (see e.g.\cite{Gl56}). The force on a cylindrical TM in
the $z$ direction produced by an axially symmetric FM can be
conveniently expressed as (see Eq.~\ref{eq:massint_main},
Sec.~\ref{sec7})

\begin{align}
&F_z = 2 M_{TM} \times \notag
\\
&\sum_{n=0}^\infty V_0^{(2n+1)}
  \sum_{i=0}^n \frac{1}{\left( -4 \right)^i}
\frac{1}{i!\left(i+1\right)!} \frac{1}{\left( 2 n- 2 i  +1
\right)!} b^{ 2 n -2 i } r^{2i} \
\end{align}
where $M_{TM}$ is the mass of a cylindrical TM with radius r and
height b, and $V_0^{(2n+1)}$ is the $2n+1$st derivative of the
gravitational potential with respect to $z$ evaluated at the
center of mass of the TM ($r=0$, $z=z_0$).

The potential $V(r=0,z)$ of the various FM components having axial
symmetry was determined analytically for three types of axially
symmetric bodies, namely a hollow ring with rectangular cross
section, one with triangular cross section and one with circular
cross section. This allows one to calculate the gravitation
potential of the tank walls and the mercury content of the tank as
a sum of such bodies. For example, the region between measured
heights on the top plate and $z=0$ at two values of the radius was
represented by a cylindrical shell composed of a right triangular
torus and a rectangular torus (i.e. a linear interpolation between
the points describing the cross section of the rings). O-rings
were calculated employing the equation for rings with circular
cross sections. A total of nearly 1200 objects (point masses and
rings of various shapes) were required to describe the two FM's.

The derivatives of the potential were evaluated using a numerical
method called ``automatic differentiation" (see e.g. \cite{Ra81}).
For the geometry of the present experiment, the terms in the
summation over $n$ decrease rapidly so that 8 terms were
sufficient for an accuracy of 0.02~ppm in the mass-integration
constant.

\subsection{Positions of TM's and FM's}
\label{sec41} 
\begin{figure}[htb]
\includegraphics[width=8.5 cm]{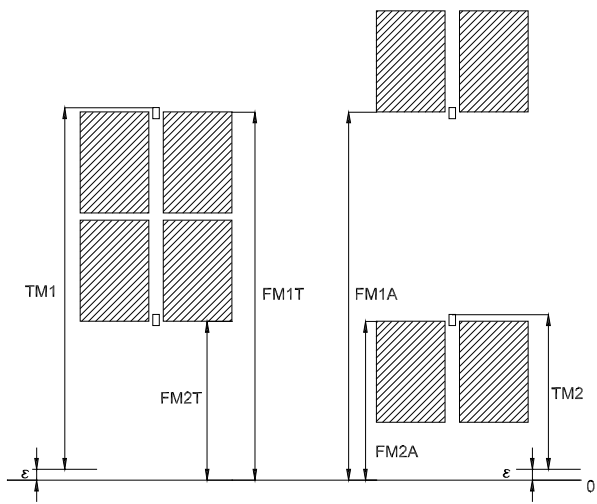}

 \caption[Geometry.]{Drawing showing the measured
vertical distances to TM and FM surfaces for the two FM positions
(T=Together and A=Apart).} \label{fig:geom}
\end{figure}

In order to carry out the mass integration, one
needs accurate weight and dimension measurements of the TM's and
FM's as well as distances defining their relative positions. The
dimension and weight measurements for TM's  were described in
Sec.~\ref{sec32}. The measurement of the TM positions shown in
Fig.~\ref{fig:geom} will now be addressed.

A special tool was made to adjust the length of the tungsten wires
under tension. Each wire made a single loop around the post on
either side of the TM and a thin tube was crimped onto the wires
to hold them together thereby forming the loop (see
Fig.~\ref{fig:cotm}). The position of the TM could only be
measured with the vacuum tube vented. The vacuum tube was removed
below a flange located at a point just above the upper TM. The TM
hanging from the balance was then viewed through the telescope of
an optical measuring device to determine its position.

The vertical position of the TM's and FM's was measured relative
to a surveyor's rod which was adjusted to be vertical. The bottom
of the surveyor's rod was positioned to just touch a special
marker mounted on the floor of the pit. The surveyor's rod had
accurate markings every cm along its length. A precision levelling
device in which the optical axis of the telescope could be
displaced by somewhat more than a cm was then used to compare the
position of the upper surface of a TM with a mark on the
surveyor's rod. Similar measurements were made for the FM's. These
measurements were made before and after each of the three
gravitational measurement(Cu, Ta 1 and Ta 2 TM's). Although the
measuring device including the surveyor's rod was removed from the
pit after each of these measurements, the reproducibility of each
position measurement was found to be better than 35 $\mu m$. The
accuracy of the average of the two sets of position measurement
for each type of TM including systematic uncertainty was 35~\mum.

A small vertical displacement of the TM's occurred when the system
was evacuated. This was measured during the evacuation of the
system by observing the TM's through the windows on the side of
the vacuum tube using the levelling device that was also used for
measuring the TM position in air. The vertical displacement was
measured several times and found to be 0.10(3)~mm. This
displacement is shown as $\varepsilon$ in Fig.~\ref{fig:geom}).

The angle of the TM axis relative to the vertical direction was
also determined with the same precision levelling instruments by
measuring the height on the top surface of the TM at two opposite
points near the outer radius of the TM. This was done for each TM
from a viewing direction almost perpendicular to the plane of the
supporting wires. The angle of the axis relative to the vertical
was found to be less than $1^\circ$ for both TM's.

The horizontal positions of the TM's were determined using a
theodolite. The left and right sides of the TM were viewed through
the telescope of the theodolite relative to an arbitrary zero
angle. The horizontal position of the central tube was determined
relative to the same zero angle. These measurements were made for
each TM and FM from two nearly perpendicular viewing angles. The
measurements were made before and after each gravitational
measurement. The radial positions of the upper and lower TM's
relative to the common axis of the FM's were found to be 0.45~mm
and 0.50~mm, respectively. The overall accuracy of the radial
positions of the TM's from these measurements was 0.1~mm. This
uncertainty results in only a small uncertainty in the value of G
determined in this experiment due to the extremum of the force
field in the radial direction. No problem was experience with
pendulum oscillation during these measurements as the amplitudes
were strongly damped in air.

\subsection{Dimensions and Weight of the FM's}
\label{sec42} The individual parts of the FM's were weighed at
METAS with an accuracy of 1~g. The weights were corrected for
buoyancy to obtain the masses.

The narrow confines of the pit made measurement of the FM's
deformed by the mercury load difficult. Although measurements of
the individual pieces before assembly were in principle more
accurate, the loading and temperature difference between dimension
measurements and gravitational measurement reduced the accuracy of
these measurements. In addition, it is known that long term
loading can release tensions in the material which result in
inelastic deformation of the material. Therefore, the measurements
made with mercury load were always used in the analysis when
available.

The uncertainty in the height of the central piece proved to be
very important in determining  the uncertainty of the
mass-integration constant. Due to the various types of
measurements for this dimension with different systematic effects,
we decided that the best value would be an equally weighted
average of the four available measurements with the uncertainty
being determined from the rms (root-mean-square) deviation from
the mean. The measurements employed were the following: (1) a
Coordinate Measuring Machine (CMM) measurement before the tanks
were assembled, (2) a Laser Tracker (LT) measurement with mercury
loading during the experiment, (3) a LT measurement in the machine
shop with no loading after completion of the experiment and (4) a
CMM measurement after the tanks had been disassembled at the end
of the experiment. The two CMM measurements were independent in
that they were made with different CMM devices and with different
temperature sensors. The uncertainty in the height as determined
from the rms deviation was 19~\mum\ for the upper tank and 9~\mum\
for the lower tank.

\begin{figure}[htb]
\includegraphics[width=8.5cm]{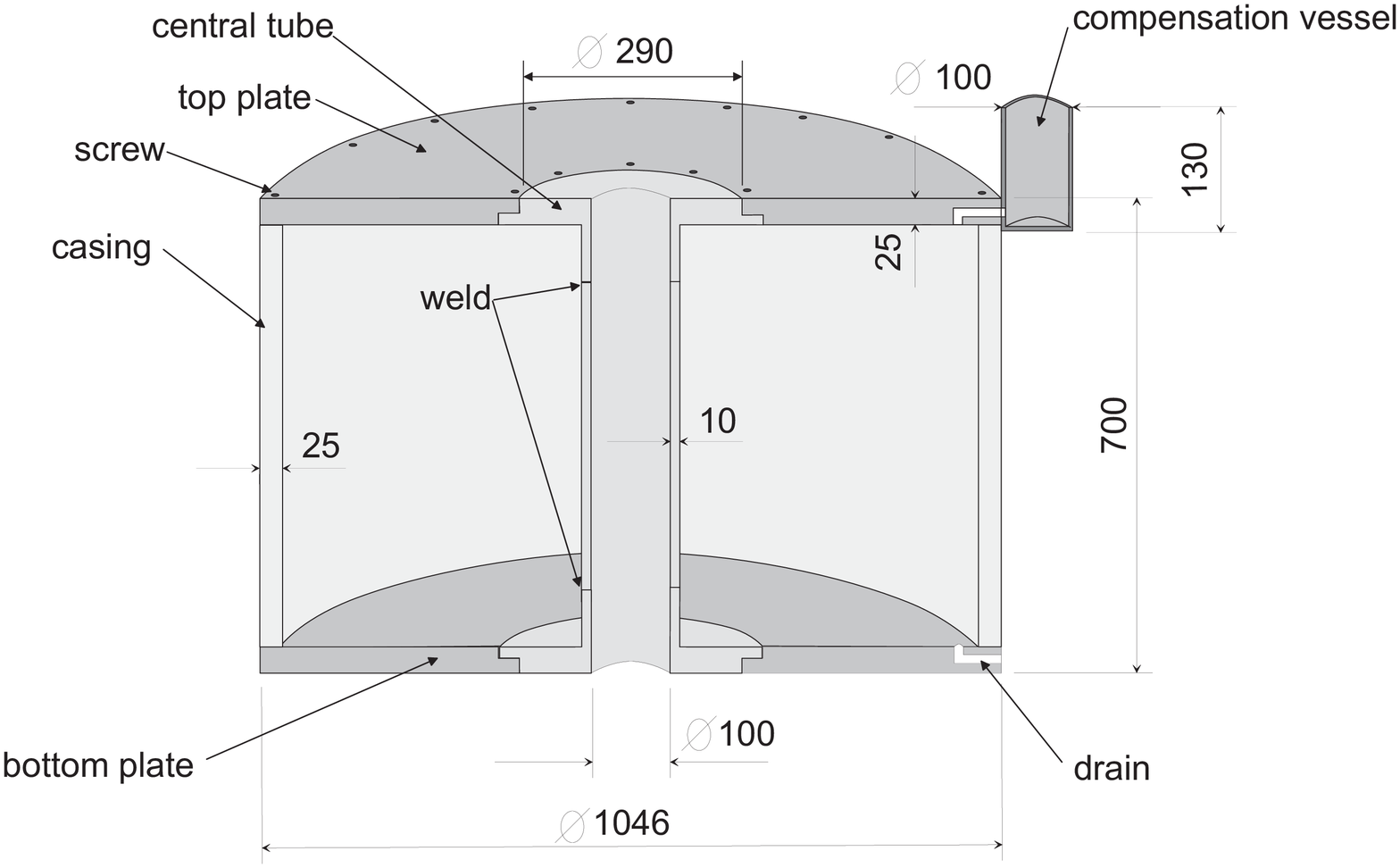}
\center{figure} \caption[Drawing of the field mass.]{Drawing of
the field mass. All dimensions are given in mm.} \label{fig:tank}
\end{figure}

A cut-away drawing of a FM tank is shown in Fig.\ref{fig:tank}. All pieces
were made from stainless steel type No.~1.4301 which is resistant
to mercury and has a low magnetic susceptibility. The pieces were
sealed to one another with mercury resistant Perbunan O-rings. The
top and bottom plates were fastened to the central piece with 12
screws. The top and bottom plates were screwed to the outer casing
with 24 screws.

Due to the nearness of the central piece to the TM's, especially
close tolerances were specified for this piece. The central piece
was annealed before machining to remove tensions which could
deform the piece during machining. A surface roughness value of
$<1$ $\mu m$ was obtained by grinding the surface after machining.
The roughness value is defined as the average height of the peaks
times the area of peaks relative to the total area of the surface.

The top and bottom plates were less critical than the central
piece. A surface roughness value of 6 to 10~\mum\ was specified
for the machining of the inner surface of these pieces. The inner
surface of the casing was even less critical and a surface
roughness value of 15~\mum\ was specified for it.

Since the mercury was filled into the tanks under vacuum
conditions, it filled more or less the exact surface profile (i.e.
the region between the grooves caused by machining). The mercury
in the filled tank was under positive pressure on all surfaces
including the top plate. The over pressure ranged from about 0.1
bar on the top plates (due to the height of mercury in the
compensation vessel shown in Fig.~\ref{fig:tank}) and 1 bar on the
bottom plates. The bulging of the central cylinder and the outer
casing due to the mercury pressure was calculated using the
equations for thin cylindrical shells \cite{Ti87}. The inward
bulging of the central tube was found to have a maximum value of
0.17~\mum\ which is completely negligible for the present
experiment. The maximum outward bulging of the outer casing was
approximately 4~\mum\ and resulted in a small (8~ppm) correction
of the volume. The loading of the tanks produced a 2~\mum\
elongation of the outer casing corresponding to a relative volume
increase of 1~ppm.

Before the tanks were assembled, measurements of the individual
pieces were made with a coordinate measuring machine (CMM).  The
inner diameters of the central tube were checked with a special
dial gauge and found to agree with the CMM values. The wall
thickness of the outer casing, top and bottom plates were measured
with a wide jaw micrometer having a dial gauge readout.

\begin{figure}
\includegraphics[width=8.5cm]{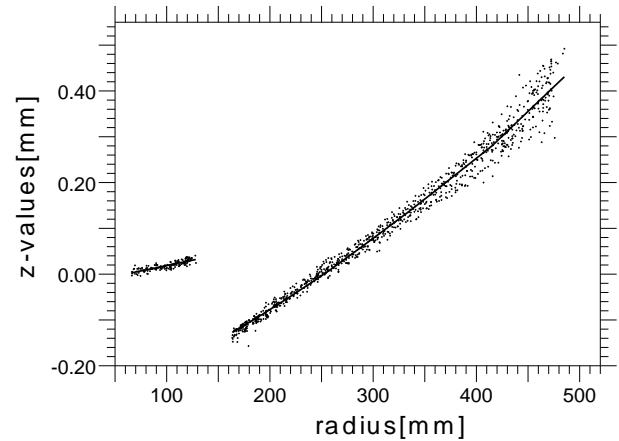}
\caption{Data and fit for the upper surface of the lower tank as a
function of radius.  The solid curve is the fit function based on
the theory of circular plates. The offset has been chosen such
that the z-value is 0~mm at a radius of 60~mm.}
\label{fig:lt}
\end{figure}

After the tanks were filled with mercury, measurements of the
outer surface of the top and bottom plates were made with a
laser-tracker (LT). It was not possible to measure the outer
surface of the casing due to the close confines of the pit. The LT
measurements were made in the dynamic mode in which the retro
reflector is moved on the surface and readings are taken as fast
as possible (1000 per s). An example of these data for the top
surface of the lower tank is shown in Fig.~\ref{fig:lt}. The data
for the other plates are similar. The force due to the mercury
loading which tended to stretch the central tube, and to a lesser
extent the outer casing, could be determined from these data using
equations based on thin axial symmetric plates and shells
\cite{Ti87}. The force on the central tube was calculated to be
($17\pm 8$~kN) which resulted in an elongation of 15~\mum.

Finally, after the mercury had been removed from the tanks,
additional LT measurements were made of the outside height of the
central tube in order to clear up a discrepancy of this dimension
as measured with the CMM and LT.

The form of the central tube and outer casing as determined by the
analysis of the measurements was very nearly circular; however,
the deviation from perfect roundness was larger than the expected
accuracy of the measurements. For the purpose of determining the
actual accuracy of the measurements, a least-squares fit was made
to the data employing a number of Fourier terms. The uncertainty
of the measured data was then determined by setting the $\chi^2$
of this fit equal to the DOF.

For the purpose of determining the uncertainty of the volume and
the mass-integration constant, we have employed effective
dimensions describing a hollow, right circular cylinder. The
effective value of the small radius $r$ and height $h$ of the
central piece were assigned the approximate values of 60 mm and
650 mm, respectively. The effective inner radius $R$ of the casing
was then determined such that the inside volume of the tank was
the value determined from measurements.

Besides the accuracy of the directly measured dimensions, one has
also to consider the effect of the expansion coefficient of the
stainless steel times the temperature difference between
measurement of the piece and the temperature during the
gravitational measurement, accuracy of the surface roughness and
the deformation due to the load. The accuracy of the temperature,
roughness and deformations effects are assumed to be one half of
the change caused by these effects. An estimate of these
accuracies for the inside effective dimensions of the tanks is
given in Table~\ref{tab:tankaccuracy}.

\begin{table}[tbp]
\caption{Estimated uncertainty in $\mu m$ of the effective values
for the radius $r$ of the central tube, the radius $R$ of the
outer casing and the height $h$ of the central tube due to various
effects. The uncertainties apply to the inside dimensions of the
tank.}
\label{tab:tankaccuracy}
\begin{ruledtabular}
\begin{tabular}{l r r r}
\multicolumn{1}{l}{Description}&
\multicolumn{1}{c}{$\sigma_r$} &
\multicolumn{1}{c}{$\sigma_R$} & \multicolumn{1}{c}{$\sigma_h$}\\
\hline measurement & 1.0 & 5 & 20
\\thickness        & - & - & 5
\\temperature      & 0.3 & 4 & 6
\\elongation       & - & - & 7
\\bulge            & 0 & 2 & -
\\roughness        & 0 & 6 & 7
\\ \hline
Total & 1.1 & 9 & 24
\\
\end{tabular}
\end{ruledtabular}
\end{table}

\subsection{Weighing the Mercury}
\label{sec43}
After a preliminary measurement in which the tanks were filled
with water, the water was drained from the tanks and they were
ventilated with warm dry air. The tanks were then evacuated to a
pressure of $10^{-2}$~mbar using a rotary pump with an oil filter
to prevent back streaming in preparation for being filled with
mercury.

Since the tanks were to be filled with mercury only once, every
effort was made to weigh the mercury as accurately as possible
during the filling. The mercury (specified purity 99.99\%) was
purchased in 395 flasks each weighing 36.5~kg and containing
34.5~kg of mercury. The general procedure for filling each tank was as
follows. The outer surface of the flasks were first cleaned with
ethanol. Half of the flasks were brought into a measuring room
near the experiment and allowed to come into equilibrium with the
room atmosphere for a few days. These flasks were then weighed
over a period of one week. Then, one after another, the flasks
were attached to the transfer device. Most of the mercury in a
flask was transferred via compressed nitrogen, first into an
intermediate vessel used as a vacuum lock and then into the
evacuated tank. A small amount of mercury was intentionally left
in each flask in order not to transfer any of the thin oxide layer
floating on the surface. The filling of a tank required one week.
After completion of the filling, the flasks with their small
remaining mercury content were weighed again.  The entire process
was then repeated for the second tank.

Various precautions and test measurements were undertaken to
insure the accuracy of the weighings. The balance employed for
these measurements was type SR 30002 made by Mettler-Toledo. The
balance was operated in the differential mode with accurately
known (5~mg) standards for weights less than 2~kg (for empty
flasks) and a stainless steel mass of approximately 36.6~kg made
in our machine shop (for full flasks). The mass of the 36.6~kg
weight was calibrated several times at METAS and remained constant
within the 18~mg (the certified accuracy of the weighings) during
the weighing of the mercury. The reproducibility of the weighing
of an almost empty flask or full flask was 20~mg. The average of
three weighings was made for empty and full flasks with the
balance output transferred directly to a computer via an RS232
interface. The balance was checked for nonlinearity and none was
found within the accuracy of the standard weights. A centering
table was used which allowed the flasks to be off center by as
much as 2~cm without influencing the measurement. Atmospheric data
used for buoyancy corrections were taken several times a day. A
10~kg calibration test was made once a day and the balance zero
was checked every hour. It was found that the flasks were
magnetized along their symmetry axis. Rotating the flask on the
balance did not change the measured weight but inverting the flask
resulted in a 100~mg difference. Since weighings were always made
with the flasks in an upright position, no magnetization error
occurred in the difference between full and empty flasks. The
variation of the weight for 12 flasks was monitored over a period
of  8 weeks. The variation was similar for all 12 flasks and
amounted to about 20~mg per flask during the 3 weeks required to
weigh the flasks and fill the tank. Mercury droplets which had not
reached the tanks and small flakes of paint which had accidentally
been removed from the outer surface of the flasks during the
transfer process were collected and weighed.

The total uncertainty in the approximately 6760~kg of mercury in
each tank was 12~g for the upper tank and 15~g for the lower tank
which gives a relative accuracy of 2.2~ppm for the mass of mercury
in the tanks. A listing of the various weighing uncertainties is
given in Table 3. An estimate of the  mercury and flask residue
that had not been collected (and the uncertainty of the amount
collected) was assumed to be 20\% of the amount collected. The
pressurizing of the flasks with nitrogen during transfer resulted
in a buoyancy correction of the almost empty flasks due to the
difference in density between air and nitrogen. The flasks were
sealed after use, but air gradually replaced the nitrogen. The
assumption of a constant density (approximate equation) for the
gas left in each empty flask during the time of the measurement
amounted to an uncertainty of 7 and 8~g for the upper and lower
tanks, respectively. The change in mass of the flasks during the
three week measurements gave an uncertainty of 4~g for each tank.
The accuracy of the standard masses caused an uncertainty of 3.7~g
for each tank.  Air buoyancy uncertainties resulted in an
uncertainty of 1.2~g for each tank. The statistical uncertainty
due to reproducibility of the balance was 0.4~g  for each tank.
\begin{table}[tbp]
\caption{A listing of the weighing uncertainties for the upper and
lower tanks. All uncertainties except that of the standard masses
are independent for each tank. The total mass of mercury in each
tank was approximately 6760 kg.}
\label{tab:mercury}
\begin{ruledtabular}
\begin{tabular}{ l  c  c }
\multicolumn{1}{ l }{Type of Uncertainty}&
\multicolumn{1}{c}{Upper}& \multicolumn{1}{c}{Lower}\\
&
\multicolumn{1}{c}{Tank[g]}& \multicolumn{1}{c}{Tank[g]}\\

\hline loss of mercury and residue from flasks & 8 & 11
\\approximate equation & 7  & 8
\\mass variation of flasks & 4 & 4
\\uncertainty of standard masses & 3.7 & 3.7
\\buoyancy correction & 1.2 & 1.2
\\ weighing statistics & 0.4 & 0.4
\\ \hline total uncertainty & 12 & 15
\end{tabular}
\end{ruledtabular}
\end{table}

\subsection{Mass Distribution of the Central Piece}
\label{sec44}
Although the principle mass making up the FM's was mercury and
therefore had only a negligible density variation due to its
pressure gradient, the density variation of the walls of the tanks
was important in determining the mass-integration constant. In
particular, the two central pieces which were very near to the
TM's and which were composed of three different pieces of
stainless steel welded together were critical for this
determination. Therefore after the gravitational measurements had
been completed, the central pieces were cut into a number of rings
in order to determine the density of these rings and thus the mass
distribution of the central pieces.

As shown in Fig.~\ref{fig:tank}, each central piece was composed
of upper and lower flanges and a central tube. The material of the
flanges extended about 40~mm beyond the surface of the flanges in
the form of the central tube. In order to determine the vertical
mass distribution in the flanges, three rings of 10~mm height were
cut from each of these 40~mm sections with the last ring
straddling the weld between flange and central tube. The weight
and dimensions of the various pieces (12 rings, 4 flanges and two
central sections of the tube) were used to determine the densities
of these pieces. The densities of the flanges were found to be
between 7.9138 to 7.9147~g cm$^{-3}$ and the densities of the
central section of tubes were 7.9062 and 7.9101~g cm$^{-3}$. The
accuracy of the absolute-density determinations was somewhat
better than 0.001~g cm$^{-3}$. The density of the weld regions did
not differ significantly from that of the flanges. The variation
of the flange densities over the 65~mm of the flange and adjoining
section of the tube was found to be less than 0.005~g cm$^{-3}$.
From these measurements, it was not possible to determine a radial
density gradient of the flanges. For calculating the effect of a
radial gradient on the mass-integration constant, we shall make
the assumption that the radial density variation was less than
0.005~g cm$^{-3}$ over the radial dimension of the flange
(160~mm). The vertical density gradients of the central tubes were
not important since their effects on the gravitational signal are
almost completely cancelled due to the symmetry of central tube
relative to the apart-together measurements with the FM's.

\subsection{Using the Measured Dimensions}
\label{sec45}
The first step in calculating the mass-integration constant was to
enter the measured dimensions and masses of the various pieces in
a computer program. For pieces that had essentially axial symmetry
such as the central tube and the outer casing, an average radius
was determined from the data measured at each height and used in
further calculations. For horizontal surfaces which were nearly
planar such as the top and bottom plates, an average height was
determined from the data at each measured radius and used in
further calculations. Since the original data were normally
available only in Cartesian coordinates, it was necessary to
determine the symmetry axis and make the conversion to cylindrical
coordinates. For data without axial symmetry such as screws, screw
holes and linear objects, single or multiple point masses were
used. With this reduced set of dimensions, approximately 580 data
elements were necessary to describe each tank.

As a preliminary calculation related to the mass-integration
constant, the volume of the individual pieces and the inside
volume of the tanks were calculated from the reduced dimensions.
Using the known weight of the piece, the calculated volume allowed
the density of the material to be determined. This was a valuable
test to check whether the input data for the piece was reasonable.

The volume for pieces with axial symmetry was determined by making
a linear interpolation between the points of the reduced data. The
volume of a piece was thus composed of a sum of cylindrical rings
with rectangular and right triangular form. A cylindrical ring
with circular cross section was used for the volume of O-rings.
The accuracy of the linear approximation in the volume
determination was checked relative to a quadratic approximation of
the surface. The linear approximation was found to be sufficient
for all calculations.

The original CMM measurements had been made for 12 $\varphi$
angles at 14 heights on the central tube, at 4 radii on the
horizontal surfaces of the central flange, at 11 radii on the
horizontal surfaces of the top and bottom plates and at 7 heights
on the outer casing. Although many more points of the horizontal
surfaces were measured with the LT, they were reduced to the
original CMM points for the purpose of volume integration and mass
integration by fitting functions to the LT data. Only outside
surfaces were measured with the LT. Inside dimensions were
obtained from the LT data by subtracting the micrometer-thickness
values. The only measurements of the  casing radius were the CMM
measurements of the inner radius. The outer surface of the casing
was determined from the CMM values combined with the micrometer
data.

\subsection{Density Constraint}
\label{sec46}
 Since the mercury represented roughly 90\% of the
total tank mass, special attention was given to its contribution
to the signal. The density of mercury samples from each tank was
measured at the Physikalisch-Technische Bundesanstalt,
Braunschweig with an accuracy of 3 ppm. One can use this density
and mass measurements of the mercury (see Sec.~\ref{sec43}) to
obtain better values for the effective tank dimensions and thus
for the mass-integration constant. This results in a correlation
among the effective parameters, $r,R,h,m$. The method employed to
determine the best parameters representing the effective values
(determined from measurements as described in the previous
section) is based on minimizing a $\chi^2$ function of the form
\begin{align}
\chi^2=&\left(\frac{r-r_0}{\sigma_r}\right)^2
+\left(\frac{R-R_0}{\sigma_R}\right)^2
+\left(\frac{h-h_0}{\sigma_h}\right)^2 \notag \\
&+\left(\frac{m-m_0}{\sigma_m}\right)^2
+\left(\frac{\rho-\rho_0}{\sigma_\rho}\right)^2
\end{align}
subject to the density constraint
\begin{equation}
\rho=\frac{m}{\pi (R^2-r^2)h}.
\end{equation}
After substituting $\rho$ from Eq. (4) in Eq. (3), $\chi^2$
becomes a function of the four parameters $r$, $R$, $h$, $m$, the
five measured quantities $r_0$, $R_0$, $h_0$, $m_0$, $\rho_0$ and
their uncertainties (see Table~\ref{tab:fitval} for the
uncertainties of the effective values). The simplex method was
used to minimize $\chi^2$ and thereby obtain best values for the
fit parameters. Although $\rho$ is not explicitly one of the fit
parameters, a best value for $\rho$ can be obtained by
substituting best fit parameters in Eq. (4).

The difference between best fit parameters and the measured values
are shown in Table~\ref{tab:fitval} along with the resulting
minimum $\chi^2$. It is seen that the difference between
parameters and measured values is less than the error in all cases
and that $\chi^2$ is consistent with the expected $\chi^2$ for a
least-squares fit with one DOF.

\begin{table}[tbp]
\caption{The correlated measured values, their uncertainties and
the difference between the best fit parameters and the measured
values for upper and lower tanks labelled 1  and 2. Only
approximate values are shown for the measured quantities.}
\label{tab:fitval}
\begin{ruledtabular}
\begin{tabular}{l r l r l r l r l}
\multicolumn{1}{l}{}&
\multicolumn{2}{c}{Measured}& \multicolumn{2}{c}{Uncertainty}&
\multicolumn{2}{c}{Difference 1}&
\multicolumn{2}{c}{Difference 2}\\
\hline $r_0$ &60   &mm   &  1.1 & $\mu \mbox{m}$ &0.007  & $\mu \mbox{m}$&0.013&$\mu \mbox{m}$
\\$R_0$      &498  &mm   &  9.0 & $\mu \mbox{m}$ &-3.5  &$\mu \mbox{m}$&-7.1&$\mu \mbox{m}$
\\$h_0$      &650  &mm   & 24.0 & $\mu \mbox{m}$ & -9.5 &$\mu \mbox{m}$&-19.0&$\mu \mbox{m}$
\\$m_0$      &6760 &kg   & 14.8 & g       & 0.3 &g&0.7&g
\\$\rho_0$   &13.54&g/cm$^{3}$       & 40.6 & $\mu \mbox{g/cm}^{3}$ &-1.3 &$\mu \mbox{g/cm}^{3}$&-2.5&$\mu \mbox{g/cm}^{3}$
\\ \hline $\chi^2$&&&&&0.27&&1.18&
\end{tabular}
\end{ruledtabular}
\end{table}

In order to obtain the uncertainty of the mass-integration signal
one needs the parameter-error matrix involving $r$, $R$, $h$, $m$
and $\rho$ multiplied by partial derivatives of the signal with
respect to the these quantities. The partial derivative with
respect to $\rho$ is zero since it does not occur explicitly in
the expression for the signal. The error of the signal $S$ is
given by
\begin{displaymath}
   \sigma_S=\left( \sum_{i,j}\frac{\partial S}{\partial x_i}
   \frac{\partial S}{\partial x_j} \textrm{err}(x_i,x_j) \right)^{1/2}
\end{displaymath}
where the $x_i$ and $x_j$ are any pair of the measured quantities
and err$(x_i,x_j)$ is the 5 by 5 parameter-error matrix. Assuming
that, for small variations about the measured values, the fit
parameters represented by the 5-dimension vector $\overline y$ can
be expressed as a linear function of the measured values
$\overline x$ of the form $\overline y=T \overline x +\overline
a$. The parameter-error matrix can be written as
\begin{displaymath}
  \textrm{err}(y_i,y_j)=T V T^t
\end{displaymath}
where T is the Jacobi derivative of $y$ with respect to $x$, $T^t$
is its transpose, $V$ is the 5 by 5 matrix covariance matrix (i.e.
$V_{i,j}=err(y_i,y_j)$) with all zero elements except along its
diagonal and $\overline a$ is a constant vector. The elements of
the matrix $T$ and the vector $\overline a$ were determined
numerically by solving a system of linear equations in which the
fit parameters were determined for measured values incremented by
small amounts ($\sigma_x$).

The partial derivatives were also determined numerically by
calculating the signal for measured values with small increments
($\sigma_x$). The signal was calculated using actual dimensions of
the deformed tanks corrected by factors relating the $r,R,h$
parameters to the effective dimensions $r_0,R_0,h_0$. The
resulting covariance matrix representing the square of the
uncertainty in the calculated mass-integration signal is shown in
Table~\ref{tab:covar}. It is seen that the elements involving $R$
and $h$ are much larger than those for $r$ and $m$. For the upper
tank (tank 1), the relative uncertainty of the calculated
mass-integration constant due to the correlated dimensions is
2.14~ppm. For tank 2, it is 2.41~ppm. The large cancellation
occurring in the sum of the elements results in the uncertainty
for these constrained parameters being approximately a factor of
seven smaller than the uncertainty that would be obtained without
the density information.

\begin{table}
\begin{ruledtabular}
\caption{The covariance matrix  involving $r,R,h,m$ for the
uncertainty of the calculated signal. Values are given in units of
ng$^2$. Only the upper part of the symmetric matrix is shown. The
sum of all element in the full matrix is 2.84~$\mbox{ng}^2$. The
sum of a similar diagonal matrix for uncorrelated $r,R,h,m$ values
(not shown) is 175 $\mbox{ng}^2$.} \label{tab:covar}

\begin{tabular}{l uuuu}
  & \mone{r}   & \mone{R}  &  \mone{h}  & \mone{m} \\ \hline
r & 0.52       &      0.04       &   0.05           &  0.0003\\
R & \mone{}    &      35.77      & -42.52           & -0.20\\
h & \mone{}    &  \mone{} &  51.62                  & -0.25\\
m & \mone{}    &  \mone{} & \mone{}             &  0.66 \\
\end{tabular}
\end{ruledtabular}
\end{table}

\subsection{The Effect of Air Density}
\label{sec47} Since air is not present in the region of the FM's,
the motions of the FM's results in a force on the TM's due to the
mass of the air elsewhere. It is as if there were a negative
contribution to the mass of the FM's due to the lack of air in
this region. This effect depends upon the density of the air in
the region surrounding the FM's.

The air pressure, the relative humidity and the air temperature
were recorded every 12~min during the gravitational measurement
thereby providing the information necessary to determine the air
density. The effect of air density on the calculated
mass-integration constant was approximately 100 ppm. The variation
of the mass-integration constant for this effect was only about 1
ppm. Thus, it was sufficient to employ only the average value of
the air density during the entire gravitational measurement. The
average density employed was 1.156~$\mbox{kg}/\mbox{m}^3$.

\subsection{The Effect of Mercury Expansion}
\label{sec48}
Due to the small temperature variations of the FM's,
the volume of the mercury relative to the volume of the tanks
changed slightly during the gravitational measurement. This
resulted in a variation of the mercury height in the compensation
vessels. The height of the mercury in each compensation vessel was
recorded every 12~min during the experiment. The calculated
mass-integration constant varied by only 0.3~ppm due to this
effect. Therefore, only an average value of the mercury height in
each compensation vessel was employed in determining the
mass-integration constant for the entire measurement.

\subsection{Uncertainties Affecting the Mass-Integration Constant}
\label{sec49}
The relative uncertainties  of the mass-integration
constant due to the various measured and estimated quantities
relating to either the upper or lower TM or to either the upper or
lower FM are given in Table~\ref{tab:umassint}. The signs of the
estimated quantities have been chosen to give the largest
uncertainty of the mass-integration constant. With the exception
of the constrained quantities $r$, $R$, $h$, and $m$, all measured
quantities of this table are independent (i.e. uncorrelated). All
estimated quantities are also independent. The total uncertainty
of the mass-integration constant due to the measured quantities
listed in Table~\ref{tab:umassint} results in a relative
statistical uncertainty of 4.89~ppm. The total uncertainty of the
mass-integration constant due to estimated quantities results in a
relative systematic uncertainty of 3.25~ppm.

\begin{table}
\caption{Mass-integration constant relative uncertainties (ppm)
associated with the measured quantities. 'Upper' and 'Lower' refer
to the upper and lower FM or TM quantities. The values in
parentheses are the uncertainty of the measured quantities. Where
two measured values are listed, the first applies to the upper
object and the second to the lower object. Quantities marked with
a $^*$ are obtained from estimated limiting values. All
uncertainties are independent except for the constrained
quantities r,R,h,m. However, these constrained quantities are
independent for the upper and lower tanks.} \label{tab:umassint}
\begin{ruledtabular}
\begin{tabular}{l ss}
\mone{Measured Quantity} & \mone{Upper} &
\mone{Lower} \\
\hline  & & \\FM Quantities & &
\\ \hspace{5 mm} $r,R,h,m$ constrained                & 1.20 & 1.09
\\ \hspace{5 mm} position=2460 or 1042 mm (35 \mum) & 2.05  &
2.99
\\ \hspace{5 mm} inner radius=50 mm (1.1 \mum)       &  0.91  & 0.91
\\ \hspace{5 mm} travel=709 mm (10 \mum)        & 0.95 & 1.22
\\ \hspace{5 mm} upper plate mass=153 kg (0.9~g)    & 0.01 & 0.01
\\ \hspace{5 mm} lower plate mass=154 kg (0.45~g)   &0.02 & 0.02
\\ \hspace{5 mm} central piece mass=46 kg (0.18~g)  & 0.03 & 0.03
\\ \hspace{5 mm} outer tube mass=412 kg (0.83~g)    & 0.01 & 0.01
\\ \hspace{5 mm} central piece no density gradient & 0.03 & 0.03
\\ \hspace{5 mm} central piece z density gradient$^*$ &<0.03&<0.03
\\ \hspace{5 mm} central piece r density gradient$^*$ &<1.1&<1.1

\\ & &
\\TM Quantities & &
\\ \hspace{5 mm} radius=23 mm (5 \mum)              & 0.57 & 0.57
\\ \hspace{5 mm} height=77 mm (5 \mum)              & 0.49 & 0.87
\\ \hspace{5 mm} position=2495, 1077 mm (35 \mum)   & 0.45 & 0.32
\\ \hspace{5 mm} mass=1.1 kg (300 \mug)             & 0.14 & 0.14
\\ \hspace{5 mm} off center=0.44 or 0.51 mm (0.1 mm)&1.03 & 1.03
\\ \hspace{5 mm} angle relative to vertical $^*$   &<1.85 &<1.85
\\ \hspace{5 mm} relative z density gradient$^*$    &<0.9&<0.7
\\ \hspace{5 mm} relative r density gradient$^*$    &<0.02&<0.02
\\ \end{tabular}
\end{ruledtabular}
\end{table}

In addition to the uncertainties related to either TM alone, there
is the common vertical displacement (shown as $\varepsilon$ in
Fig.~\ref{fig:geom}) of both TM's due to evacuating the system.
The uncertainty of this displacement results in a relative
uncertainty of the mass-integration constant equal to 0.78~ppm
which is added to the other uncertainties as an independent
relative uncertainty. Including the uncertainty of $\varepsilon$
results in a relative statistical uncertainty of the
mass-integration constant equal to 4.95~ppm.

\section{Discussion of Measurements}
\label{sec5} The measured gravitational signal discussed in
Secs.~\ref{sec37} to \ref{sec311} is 784,883.3(12.2)(5.8)~ng. The
calculated mass-integration constant determined in
Secs.~\ref{sec41} to \ref{sec45} is 784,687.8(3.9)(2.6)~ng. Using
these values, we obtain the value for the gravitational constant
\\ \\
$G$=6.674252(109)(54)$\times 10^{-11}\;\Gunit$.
\\ \\
A summary of the relative uncertainties contributing to this
result is given in Table~\ref{tab:relerrs}.

\begin{table}
\caption{Relative statistical and systematic uncertainties of $G$
as determined in this experiment.} \label{tab:relerrs}
\begin{ruledtabular}
\begin{tabular}{l l r}
{Description}& {Statistical(ppm)} &{Systematic(ppm)}
\\ \hline
Measured Signal & &\\
\hspace{5 mm} Weighings & 11.6 & \\
\hspace{5 mm} TM-sorption  & 7.4 & 7.4\\
\hspace{5 mm} Linearity  &6.1& \\
\hspace{5 mm} Calibration & 4.0 & 0.5\\
Mass Integration & 5.0&3.3\\
\hline
Total &  16.3 & 8.1 \\
\end{tabular}
\end{ruledtabular}
\end{table}
The relative statistical and systematic uncertainties of this
result are 16.3~ppm and 8.1~ppm, respectively. The two largest
contributions to the total relative  uncertainty are the
statistical uncertainty of the weighings (11.6~ppm) and the
combined statistical and systematic uncertainty due to the
TM-sorption effect (10.3~ppm). All uncertainties have been given
as one sigma uncertainties. Statistical and systematic
uncertainties have been combined to give a total uncertainty by
taking the square root of the sum of their squares.

\subsection{Comparison with Our Previous Analysis}
\label{51}
Our previously published value \cite{Sa02} for $G$ was
$6.674070(220) \times 10^{-11}\; \Gunit$ which was based on the
measurements of both the copper and tantalum TM's. The value for
the copper TM's alone was $6.674040(210)\times 10^{-11}\;\Gunit$.
The value obtained for $G$ in the present analysis for only the
copper TM's ($6.674252((124)\times 10^{-11}\; \Gunit$) is in
reasonable agreement with the previous value. The difference
between the present and previous result is due primarily to the
correction for the ZP curvature which was not taken into account
in the previous analysis. A minor difference is also due to a
slightly different selection of the analyzed data.

The uncertainty given for the present result is appreciably
smaller than that of our previous result. This is due to a better
method used in computing the nonlinearity correction
(Sec.~\ref{sec310}) and a  calculation of the mass-integration
constant (Sec.~\ref{sec45}) using the mercury density as a
constraint. The uncertainty of the linearity correction was
reduced from 18~ppm to 6.1~ppm and the uncertainty of the
mass-integration constant was reduced from 20.6~ppm to 6.7~ppm.
The statistical uncertainty of the weighings in the present
analysis is somewhat larger than the previous value (9.1~ng vs
5.4~ng). This is due to the correlation of the ZP-corrected data
of the present analysis. The previous ABA analysis involved only
uncorrelated data.

\subsection{Comparison with Other Measurements}
\label{sec52}
\begin{figure}[width=8cm]
\includegraphics[width=8.5cm]{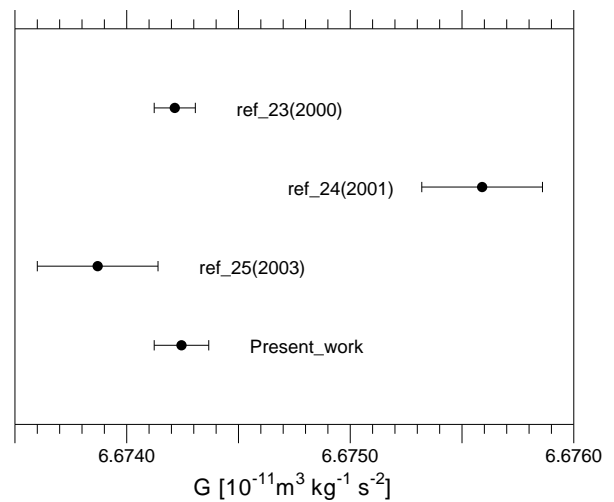}
\caption{Plot of recent measurements with relative errors less
than 50~ppm} \label{fig:Gcomp}
\end{figure}

Recent measurements \cite{Gu00,Qu01,Ar03} of the
gravitational constant with relative errors less than 50~ppm are
listed in Table~\ref{tab:Gcomp}\ and shown in
Fig.~\ref{fig:Gcomp}. We list only the latest publication of each
group. It is seen that the present result is in good agreement
with those of Gundlach and Merkowitz \cite{Gu00} and Fitzgerald
\cite{Ar03}. It is in disagreement (3.6 times the sum of the
uncertainties) with the result of Quinn et al. \cite{Qu01}.

All of the measurements listed in Table~\ref{tab:Gcomp} with the
exception of our own were performed using torsion balances. It is
therefore instructive to compare the problems encountered in the
different types of measurements.

In the measurements being discussed, the statistical accuracy in
determining the gravitational signal was obtained in measurements
lasting one to six weeks. However, as in the case of all precision
measurements, the time required for obtaining a good statistical
accuracy of the measurement is less than the time required to
obtain calibration accuracy of the equipment and the time
necessary to investigate and eliminate systematic errors.  All of
the measurements listed in Table~\ref{tab:Gcomp} have been long, on-going
investigations which have lasted for periods up to ten years.

Although the beam-balance measurement was made with more massive
FM's (15~t) than were employed in the torsion-balance measurements
($<60$~kg), the larger gravitational signal had to be measured in
the presence of the TM weight. In our experiment the gravitational
signal was roughly 0.7~ppm of the total weight on the balance.
This small ratio of signal to total weight on the balance resulted
in larger effects of zero-point drift as well as larger
statistical noise in the beam-balance data than in the
torsion-balance data. In the torsion-balance measurements, the
deflection of the balance arm is due entirely to the gravitational
force to be determined with only small perturbations from distant
moving objects.

A similar problem has to do with the change of TM weight that is
produced by an adsorbed water layer. In our measurement, this
varied with the temperature of the vacuum tube produced by the FM
motion. This resulted in one of the largest contribution to the
uncertainty of the gravitational signal (see Sec.~\ref{sec310}).
In the beam-balance measurement the weight change adds directly to
the signal whereas in the torsion-balance it adds only to the mass
of the torsion-arm and is therefore a negligible effect.

The calibration of the beam balance, while simple in principle, is
difficult in practice due to the lack of calibration masses having
the required mass and accuracy. We have used an averaging method
to correct for the nonlinearity of the balance (see
Sec.~\ref{sec311}) involving a large number of auxiliary masses.
This allowed the comparison of the gravitational signal with a
heavier, accurately known calibration mass. The accurate
calibration of the torsion balance also presents a problem.
Various methods involving either electric forces or an angular
acceleration of the measuring table to compensate the
gravitational force have been used.

In our measurement, the TM was positioned at a double extremum of
the force field produced by the FM's. This greatly reduces the
relative accuracy of the distance measurements required to
determine the mass-integration constant. It also reduces the
problem resulting from a density gradient in the TM. It is
difficult to compare the problems involved in determining the
mass-integration constant for the two types of experiments. It
appears that the distance between the field masses attracting the
small mass of the torsion balance must be measured with very high
absolute accuracy (1~\mum\  is the accuracy given for this
distance in the torsion balance experiments).

The use of liquid mercury as the principle component of the FM's
reduces the problem associated with the density gradients of the
FM's. There is still the density gradient of the vessel walls
which has to be considered. The field masses employed in the
torsion-balance experiments were either spheres or cylinders. The
FM's were rotated between measurements in order to compensate
gradient effects.

The large mercury mass resulted in deformations of the vessel
which had to be accurately determined. The FM's although, nearly
cylindrical in form, required more than 1000 ring and point-mass
elements in order to determine the mass-integration constant. A
similar problem (but on a smaller scale) occurs in the torsion
experiments in accounting for small imperfections of the spheres,
cylinders or plates and in determining their relative positions.

The determination of $G$ using a beam balance is beset with a
number of problems which we have tried to describe in detail. We
have been able to reduce the uncertainty in $G$ resulting from
these problems to values comparable to the statistical
reproducibility of the weighings determining the gravitational
signal. The total uncertainty for $G$ which we obtain with a beam
balance is comparable to the uncertainty quoted in recent
torsion-balance determinations of $G$. We believe that the
beam-balance measurement involving a number of quite different
problems than encountered in torsion balance measurements can
therefore provide a useful contribution to the accuracy of the
gravitational constant.

\begin{table}[tbp]
\caption{Recent measurements of the gravitational constant.
Statistical and systematic uncertainties have been added as if the
were independent quantities.} \label{tab:Gcomp}
\begin{ruledtabular}
\begin{tabular}{l  l}
\multicolumn{1}{l}{Reference}&
\multicolumn{1}{c}{$G[10^{-11}\;\Gunit]$}\\
\hline
Gundlach and Merkowitz \cite{Gu00} &6.674215( 92)\\
Quinn et al.\cite{Qu01} &6.675590(270)\\
Armstrong and Fitzgerald \cite{Ar03}&6.673870(270)\\
Present analysis &6.674252(124) \\
\end{tabular}
\end{ruledtabular}
\end{table}

\section{Acknowledgements}
\label{sec6}
The present research has been generously supported by
the Swiss National Fund, the Kanton of Z\"{u}rich, and
assistantship grants provided by Dr. Tomalla Foundation  for which
we are very grateful. The experiment could not have been carried
out without the close support of the Mettler-Toledo company which
donated the balance for this measurement and made their laboratory
available for our use. We are also extremely grateful to the Paul
Scherrer Institut for providing a suitable measuring room and the
help of their staff in determining the geometry of our experiment.
We also wish to thank the Swiss Metrological Institute and the
Physikalisch-Technische Bundesanstalt, Braunschweig, Germany for
making a number of certified precision measurements for us. Our
heartfelt thanks are also extended to the staff of our machine
shop for their advice and precision work in producing many of the
pieces for this experiment. We wish to thank E. Klingel\'{e} for
determining the value of local gravity at the measuring site. We
are also indebted to the firms Almaden Co., Metrotec A.G. and
Reishauer A.G. for the services they provided.

\section{Appendix}
\label{sec7} The general idea in determining the $z$ component of
force on a small volume element of a TM is first to calculate the
potential along the $z$ axis due to the FM. The off-axis potential
can then be obtained by making a Taylor series expansion for small
$r$ and substituting this in the Laplace equation. This is the
procedure which is often used in electrostatic calculations
\cite{Gl56}. The force in the $z$ direction is just the negative
derivative of this potential with respect to $z$. We present first
a derivation of the force for the axially symmetric case and then
describe the modification required for a nonaxially symmetric
potential.

The gravitational potential on the $z$ axis of a homogeneous,
torus of rectangular cross section with density $\rho_{FM}$, inner
radius $R_1$, outer radius $R_2$ and half height $Z$ is given by
the equation
\begin{equation}
  \Phi(r=0,z) = 2\pi\rho_{FM} G   \int_{-Z/2}^{Z/2} \int_{R1}^{R2}
   \frac{r' dr' \;dz'}{\sqrt{{r'}^2 + (z'- z)^2}}
\label{eq:Phi}
\end{equation}
where $r$ and $z$ are the radial and axial coordinates of a point
within the TM expressed in cylindrical polar coordinates. For
convenience, one chooses the zero of the potential at the center
of mass of a TM. For simplicity, the cross section of the FM
employed in Eq.~\ref{eq:Phi} has been chosen to be a rectangle.
Besides the torus with rectangular cross section, analytic
expressions for the two dimensional integrals with triangular and
circular cross sections were also employed.

The potential at points close to the axis can be calculated as a
power series in $r$
\begin{equation}
\Phi(r,z) = a_0 + a_1 r + a_2 r^2 + a_3 r^3 + \ldots =
\sum_{i=0}^\infty a_ir^i  \;
\label{eq:series}
\end{equation}
with unknown coefficients $a_1$,$a_2$,$\ldots$ which are functions
of $z$ alone. For $r$=0, one has $a_0=\Phi(0,z)$. The
gravitational field satisfies the Laplace equation  $\nabla^2 \Phi
= 0$. Applying the Laplace operator in cylindrical coordinates to
Eq.~\ref{eq:series} leads to
\begin{displaymath}
\nabla^2 \Phi=a_1r^{-1} + \sum_{i=0}^\infty r^i \left( \frac{d^2
a_i}{d z^2} + \left(i+2 \right)^2 a_{i+2} \right) = 0 \; .
\end{displaymath}
This equation is valid for all $r$. Since $\nabla^2 \Phi=0$ for
$r=0$, $a_1$ must be identically zero. Thus, the values for $a_i$
can be calculated recursively from
\begin{displaymath}
a_{i+2} = - \frac{1}{\left( i+2 \right)^2} \frac{d^{2}
a_i}{dz^{2}}
\end{displaymath}
starting with either $a_0$ or $a_1$. Since $a_1=0$, all terms with
odd $i$ are zero and $a_i$ for even $i$ can be obtained starting
with $a_0$. By induction, it is easily shown that the coefficient
$a_{2n}$ is
\begin{displaymath}
a_{2n} = \left( - \frac{1}{4} \right)^n \frac{1}{\left( n!
\right)^2} V^{(2n)}(z)
\end{displaymath}
with
\begin{displaymath}
   V^{(2n)}(z)=\frac{ d^{2n} \Phi(0,z) }{ d z^{2n} } .
\end{displaymath}
Here, $V^{(0)}$ is just $\Phi(0,z)$ which can be easily calculated
using  Eq.~\ref{eq:Phi}.

Using this expression for the coefficients in the
expansion~\ref{eq:series},
 the   gravitational potential can be calculated  from the sum
\begin{equation}
\Phi(r,z) = \sum_{i=0}^\infty \left( - \frac{1}{4} \right)^i
\frac{1}{\left( i! \right)^2}   V^{(2i)}(z)\; r^{2i}
\label{eq:potential_sum} \;.
\end{equation}

The gravitational field $g_z$ in z-direction is given by
$-\partial \Phi / \partial z$ or
\begin{displaymath}
g_z(r,z)  = -\sum_{i=0}^\infty \left( - \frac{1}{4} \right)^i
\frac{1}{\left( i! \right)^2}   V^{(2i+1)}(z) \; r^{2i} \; .
\end{displaymath}
Integrating $g_z$ over the volume of the TM, one obtains the force
on the TM in the $z$ direction
\begin{align}
&F_z =-2 \pi \rho_{TM} \times \notag \\
&\sum_{i=0}^\infty \left( - \frac{1}{4}
\right)^i \frac{1}{\left( i! \right)^2} \int_{-b}^{+b} \ d z'
\int_0^r   V^{(2i+1)}(z) \; r'^{2i}  r' \ d r' \;
\label{eq:F_sum}
\end{align}
where the origin has been taken to be the center of the TM. The
integration over $r$ is trivial. The integration over $z$ is
\begin{displaymath}
\int_{-b}^{+b} V^{(2i+1)} d z' =  V^{(2i)}(b) - V^{(2i)}(-b)  \;
.\nonumber
\end{displaymath}

Making a Taylor expansion for small $b$ on the right side of
equation, one obtains
\begin{eqnarray}
\lefteqn{\int_{-b}^{+b} V^{(2i+1)} d z' =}  \nonumber \\
&  V^{(2i)}(0) + b V^{(2i+1)}(0) + \frac{1}{2!} b^2 V^{(2i+2)}(0)
+ \ldots
  \nonumber \\
&    -V^{(2i)}(0) + b V^{(2i+1)}(0) - \frac{1}{2!} b^2
V^{(2i+2)}(0) + \ldots  \nonumber \; .
\end{eqnarray}
Adding similar terms results in
\begin{align*}
\int_{-b}^{+b} V^{(2i+1)} d z' = &   2 b V^{(2i+1)}(0) +
\frac{2}{3!} b^3 V^{(2i+3)}(0)\\ \notag
& + \frac{2}{5!}b^5 V^{(2i+5)}(0) + \ldots  \notag
\end{align*}
or
\begin{equation}
\int_{-b}^{+b} V^{(2i+1)} d z' =\sum_{j=0}^\infty
\frac{2}{(2j+1)!} b^{2j+1} V^{(2i+2j+1)}(0)  \; .
\label{eq:integral_sum}
\end{equation}

The final equation for the gravitational force on a cylinder can
then be calculated by combining Eq.~\ref{eq:F_sum} and
\ref{eq:integral_sum} to obtain
\begin{align}
& F_z=-2 \pi b r^2 \rho_{TM}  \notag \sum_{n=0}^\infty V^{(2n+1)}(0) \times \\
&
\sum_{i=0}^n \frac{1}{\left( -4 \right)^i}
\frac{1}{i!\left(i+1\right)!} \frac{1}{\left( 2 n- 2 i  +1
\right)!} b^{ 2 n -2 i } r^{2i} \ .
\label{eq:massint_main}
\end{align}

Only odd derivatives of the potential are required for this case
involving complete axial symmetry. The convergence of the series
can be improved by dividing the TM into two or more shorter
cylinders. The algebraic expressions for the $V^{2n}(0)$, as
determined using "automatic differentiation" \cite{Ra81}, are very
long and will not be given here.

The above derivation has assumed that the TM and FM have a common
axis of cylindrical symmetry. We will now show how essentially the
same equations can be employed for an arbitrary FM potential. This
allows one to calculate the potential of a FM with cylindrical
symmetry but with its axis displaced and/or tilted relative to
that of the TM.

Again, in order to facilitate integration over the volume of the
cylindrical TM, one employs cylindrical polar coordinates with the
$z$ axis along the symmetry axis of the TM.  The potential is now
a potential of the form $\psi(x,y,z)=\psi\left (r\cos(\varphi),
r\sin(\varphi),z\right)$ which satisfies $\nabla^2\psi=0$. The
center of mass of the TM is chosen as the zero of potential.

One defines a function $\Psi(r,z)$ such that
\begin{equation}
 \Psi(r,z)=(2 \pi)^{-1}\int_{\varphi=0}^{2\pi} \psi(r,\varphi,z)
  d\varphi.
 \label{eq:Psi}
\end{equation}
One can then show that this function satisfies the same assumption
that were made for the function $\Phi(r,z)$, namely that $\Psi$
has axial symmetry, that it is zero on the $z$ axis and that its
Laplacian is zero. It can therefore be used in the Eqs. 7 through
10 instead of $\Phi$ to obtain the gravitational force integrated
over the TM.

The axial symmetry of $\Psi$ is obvious since $\varphi$ has been
removed by the integration over $\varphi$. The zero potential was
chosen to be at the TM center of mass. The Laplacian of $\Psi$ can
be shown to be zero by allowing the Laplacian to operate on $\Psi$
as defined in Eq.~\ref{eq:Psi}. Reversing the integration and
differentiation operations one obtains
\begin{displaymath}
  \nabla^2 \Psi(r,z) = (2 \pi)^{-1} \int_{\varphi=0}^{2
  \pi} (\nabla^2 \psi-r^{-2}\frac{\partial^2 \psi}{\partial \varphi^2})
  d\varphi.
\end{displaymath}
The Laplacian of $\psi(r,\varphi,z)$ is zero for an inverse $r$
potential. The integral of the second term is $r^{-2}\partial \psi
/ \partial \varphi$ evaluated at $\varphi=0$ and $2\pi$ which is
also zero. Thus, one obtains the desired result that $\nabla^2
\Psi(r,z)=0$.

In order to use this property of a potential which does not have
axial symmetry about the TM axis (the $z$ axis), one must
determine the potential and its derivative with respect to $z$
along the $z$ axis. This is not difficult for the case of a FM
which has axial symmetry about an axis not coincident with the TM
axis. One merely uses Eq. 7 to determine the potential at radial
distances from the FM axis corresponding to points on the TM axis.
The force in the $z$ direction (TM axis) is then determined as
before using Eqs. 8 and 10. This procedure is particularly useful
for the case of TM and FM axes which are parallel but which are
displaced relative to one another.

In principle, one can determine the derivatives with respect to
$r$ which are required for the force on a TM tilted relative to
the vertical; however, in this case it is simpler to approximate
the TM by a number of thin slabs displaced from the vertical axis.
This completes the discussion of nonaxial-symmetric potentials.

\end{document}